\pdfoutput=1

\documentclass[12pt]{article}
\usepackage{epsf}
\usepackage{amsmath}
\usepackage{graphics}
\usepackage{cite}

\renewcommand{\thefootnote}{\#\arabic{footnote}}

\newcommand{\lesssim}{ \mathop{}_{\textstyle \sim}^{\textstyle <} }

\begin{document}
\setcounter{footnote}{0}
\begin{titlepage}

\hfill January 2008\\    

\begin{center}

\vskip .5in

{\Large \bf
The Hubble Constant and Dark Energy \\
from Cosmological Distance Measures
}

\vskip .45in

{\large
Kazuhide Ichikawa$^1$ and Tomo Takahashi$^2$
}

\vskip .45in

{\em
$^1$Institute for Cosmic Ray Research, 
University of Tokyo, Kashiwa 277-8582, Japan\\
$^2$
Department of Physics, Saga University, Saga 840-8502, Japan \\
}

\end{center}

\vskip .4in

\begin{abstract}

  We study how the determination of the Hubble constant from cosmological distance measures is affected by
  models of dark energy and vice versa. For this purpose, constraints
  on the Hubble constant and dark energy are investigated using the
  cosmological observations of cosmic microwave background,
  baryon acoustic oscillations and type Ia suprenovae.
  When one investigates dark energy, the Hubble constant is often a
  nuisance parameter, thus it is usually marginalized over.  On the
  other hand, when one focuses on the Hubble constant, simple dark
  energy models such as a cosmological constant and a constant
  equation of state are usually assumed.  Since we do not know the nature of
  dark energy yet, it is interesting to investigate the Hubble
  constant assuming some types of dark energy and see to what extent
  the constraint on the Hubble constant is affected by the assumption
  concerning dark energy.  We show that the constraint on the Hubble constant
  is not affected much by the assumption for dark energy.  We
  furthermore show that this holds true even if we remove the
  assumption that the universe is flat.  We also discuss how the prior
  on the Hubble constant affects the constraints on dark energy and/or
  the curvature of the universe.

\end{abstract}
\end{titlepage}

\renewcommand{\thepage}{\arabic{page}}
\setcounter{page}{1}
\renewcommand{\thefootnote}{\#\arabic{footnote}}

\section{Introduction}

Recent precise cosmological observations provide us with a large
amount of data to probe the evolution and the present state of the
universe.  We now customarily extract the information from them by
constraining cosmological parameters.  The Hubble constant, the expansion rate of the universe at present, is one of the
most important cosmological parameters and measured in various ways.

In most studies which consider the constraint on the Hubble constant
from cosmological observations,
a cosmological constant is often assumed as dark energy. However, the
nature of dark energy is not understood yet, which is one of the
challenging problem in cosmology today, thus the constraints on the
Hubble constant should also be considered assuming some possible dark
energy models other than a cosmological constant. Since the energy
density of dark energy can change some distance measures which have
been used to constrain the Hubble constant, the assumption concerning
dark energy is expected to have an influence on the constraint on it.
In some works, the constraint on the Hubble constant is obtained
without assuming a cosmological constant
(e.g. Refs.~\cite{Spergel:2006hy,Tegmark:2006az}) but a constant
equation of state is still assumed. However, a lot of models of dark
energy proposed so far have a time-varying equation of state and most
of recent works of dark energy accommodate its time dependence in some
way.  Thus, in light of these considerations and the importance of the
Hubble constant, it is interesting to study constraints on the Hubble
constant including time-evolving dark energy equations of state. This is one of the
issues which we are going to investigate in this paper.

On the other hand, notice that the Hubble constant is often treated as
a nuisance parameter over which we marginalize when we investigate
dark energy since the parameters for dark energy themselves are those
of interest in such a case. However, it should be mentioned that, when
one studies the nature of dark energy, observations of distance
measures such as the angular diameter distance which is relevant to
the position of acoustic peaks in cosmic microwave background (CMB)
power spectrum and the scale of baryon acoustic oscillations (BAO)
are often used.  Since the Hubble constant affects such
distance measures, it is expected that the prior on the
Hubble constant can affect constraints on the nature of dark energy.
Furthermore, the Hubble constant can also be determined with the cosmic
distance ladder measurements independently from the cosmological
distance measurements mentioned above. Hence investigating dark energy
with some priors on the Hubble constant is also an interesting
subject, which is discussed in this paper too.

To study the issues mentioned above, we study the constraints on the
Hubble constant and dark energy using the observations of CMB, BAO and
type Ia supernovae (SN).  For dark energy, we assume some types of
time-varying equation of state as well as the case with a constant
equation of state.  By assuming several priors on the Hubble constant
and dark energy equation of state, we can see how the determination of
one of them can affect that of the other.  In addition, we study them
allowing a non-flat universe to make our analysis more general.

The structure of this paper is as follows. In the next section, we
summarize our method for obtaining constraints from cosmological
observations.  Some parametrizations of dark energy adopted in this
paper are also briefly explained.  The data used for the analysis are
mentioned there too.  In Sec.~\ref{sec:result}, we present our results
and discuss some implications of our results for the study of the
Hubble constant, dark energy and the curvature of the universe.  In the final section, we summarize
our results and give the conclusion.  In the appendix, we give some
quantitative explanations how the combinations of CMB, BAO and SN can
break degeneracies among the Hubble constant, matter density, dark
energy parameters and the curvature of the universe.

\section{Dark Energy Parametrizations and Method} \label{sec:method}

In this section, we explain the method to constrain the cosmological
parameters such as the Hubble constant and dark energy equation of
state.  We use the data from CMB, BAO and SN. Before we describe the
method in detail, first we mention the dark energy parametrizations
adopted in this paper, which accommodate the time variations of its
equation of state.  Among various possible parametrizations of dark
energy equation of state, we use simple and often used ones.  The
first parametrization adopted in the following analysis is
\cite{Chevallier:2000qy,Linder:2005nh}:
\begin{equation}
\label{eq:eos}
w_X = w_0 + (1 -a) w_1 = w_0 + \frac{z}{1+z} w_1, 
\end{equation}
which varies in proportion to the scale factor $a$ (normalized as $a=1$ at the present epoch).
  If we define dark
energy as a fluid which can accelerate the universe, this can motivate
the following prior:
\begin{equation}
\label{eq:eos_prior}
w_X \le -\frac{1}{3}.
\end{equation}
For the particular parametrization Eq.~(\ref{eq:eos}), this is ensured
by forcing $w_0$ and $w_1$ to satisfy
\begin{equation}
\label{eq:eos_prior2}
w_0 + w_1 \le -\frac{1}{3}, ~~~ w_0 \le -\frac{1}{3}.
\end{equation}
In principle, a dark energy fluid can have its equation of state larger
than $-1/3$.  Thus a weaker prior $w_X \le 0$ is sometimes adopted,
requiring that dark energy does not dominate the universe at early
time, say at the epoch of recombination, since such early domination
of dark energy is apparently inconsistent with cosmological data.
However this prior allows the case with $w_X \sim 0$ where dark energy
behaves in almost the same way as dark matter as far as the background
evolution is concerned.  In such a case, dark energy can be considered
as a part of dark matter, which complicates the interpretation of the
constraints on dark energy \footnote{
  This holds true when one considers the background evolution alone,
  which is the case of the present paper.  If we include the
  information of perturbation of dark energy, the effect of dark
  energy can be distinguishable even if its equation of state is
  almost the same as that of dark matter.  
  In that case, we need a full CMB
  angular power spectrum analysis to obtain the constraint from CMB but such analysis is beyond the scope of this paper.
}.  With the prior Eq.~\eqref{eq:eos_prior}, the dark energy cannot be
a dominant component of the universe at early times and even its
fraction to the total energy density is negligible.  Thus we can
safely avoid the complication mentioned above by using the prior
Eq.~\eqref{eq:eos_prior}.
We provide more detailed discussion on the effects of dark energy perturbation on the CMB
power spectrum in Appendix \ref{sec:DE_pert}.

We also adopt another type of dark energy parametrization:
\begin{equation}
w_X(z) = 
\begin{cases}
 \tilde{w}_0 + \displaystyle\frac{\tilde{w}_1 - \tilde{w}_0}{z_*} z 
 &( {\rm for}~~ z \le z_*)  \\ \\
 \tilde{w}_1      & ({\rm for}~~ z \ge z_*),   
\end{cases} 
\label{eq:eos2}
\end{equation}
where we interpolate the value of $w_X$ linearly with respect to the
redshift $z$ from the present epoch back to some transition redshift
$z_*$. $w_X$ becomes $\tilde{w}_0$ at $z=0$ and $\tilde{w}_1$ for $z
\ge z_*$.  The case with $\tilde{w}_0= \tilde{w}_1 = -1$ corresponds
to a cosmological constant.  The prior Eq.~\eqref{eq:eos_prior} is
satisfied by setting $\tilde{w}_0 \le -1/3$ and $\tilde{w}_1 \le -
1/3$.  This parametrization is essentially the same as the one which
has a linear dependence on the redshift $z$ such as $w_X = w_a +
w_b\, z$. In fact, this form is adopted in many literatures with a
cutoff at some redshift to avoid large value of $w_X$ at $z \sim 1000$
which is relevant to the CMB constraint
\cite{Huterer:2000mj,Weller:2001gf,Frampton:2002vv}.  In the
following, we consider several values of $z_*$ to see how the dark
energy parametrization affects the constraints on the Hubble constant
and vice versa.
We note that this parametrization assumes a finite derivative of $w_X$ at the present epoch
so we cannot probe the dark energy model which behaves like a cosmological constant
(or a constant $w_X$) from some earlier epoch to the present but variable before that epoch.
Also, we should note that although this parametrization appears to have three parameters, 
we regard it as a family of two-parameter models labeled by $z_*$. Namely, we consider
several two-parameter models each with a different value of $z_*$. Such a distinction among 
dark energy parameters may be artificial, but it is simpler and sufficient for our goal to 
show how priors for the Hubble constant and ways of parametrizing the dark energy equation of state
(and the assumption for the curvature of the universe) could influence a conclusion drawn from 
the analysis of CMB, BAO and SN data. 

Now we give detailed descriptions of the method how we make use of
observational data. Incidentally, the model parameter fitting to the
data is performed by $\chi^2$ minimization. We use the Brent method
\cite{brent} extended to multi-parameters, as described in
Ref.~\cite{Ichikawa:2004zi}, to search a minimum efficiently.

\subsection{CMB}
To fit a model to the CMB data, 
we first note that we here do not use the whole information of the CMB power spectrum
but use only the acoustic scale $\theta_A$ and the matter density $\omega_m$ as Eq.~\eqref{eq:chi2_CMB} \footnote{
  Recently, Refs.~\cite{Elgaroy:2007bv,Wang:2007mz} noted that it is
  better to use $\ell_a = \pi / \theta_A$ and $R = \sqrt{\Omega_m
    H_0^2} r_\theta(z_{\rm rec})$ simultaneously to obtain the
  constraint from CMB.  We in this paper use rather conventional way
  of using $\theta_A$ and $\omega_m$ as in, for example,
  Refs.~\cite{Page:2003fa,Wright:2007vr}, but both ways make use
  of the same features of the CMB power spectrum, the peak position
  and height, and should give similar result. We checked that the
  constraints on the $\Omega_m$--$h$ plane from these two methods are
  almost same for the case with a cosmological constant.  
}. 
  This is because if we include the
  effects of perturbation of dark energy and perform a full CMB
  angular power spectrum analysis to obtain the constraint from CMB,
  it would be very time-consuming.  One of the purposes of this paper is
  to investigate cosmological constraints assuming various dark energy
  parametrizations. Thus, in this case, much faster method is
  preferable and it is known that if dark energy dominates the
  universe only at late time, the constraint derived from the
  information on the background evolution well captures the nature of
  dark energy. This condition is satisfied by adopting the prior Eq.~\eqref{eq:eos_prior}.

The acoustic scale $\theta_A$ which defines the characteristic angular scale of the
acoustic oscillations is written as
\begin{eqnarray}
\theta_A 
 =  \frac{r_s(z_{\rm rec})}{r_\theta(z_{\rm rec})}.
\end{eqnarray}
$\theta_A$ is given once we determine the comoving angular
diameter distance to the last scattering surface $r_\theta(z_{\rm
  rec})$ and the sound horizon at the recombination epoch $r_s (z_{\rm
  rec})$ where $z_{\rm rec}$ is the redshift of the epoch of
recombination.  The comoving angular diameter distance to the last
scattering surface is given as
\begin{equation}
r_\theta(z_{\rm rec})  = 
\frac{1}{H_0\sqrt{|\Omega_k|} }
\mathcal{S}  
\left( \sqrt{|\Omega_k|} \int_0^{z_{\rm rec}} \frac{dz'}{H(z')/H_0} \right),
\label{eq:r_theta}
\end{equation}
where $\mathcal{S}$ is defined as $\mathcal{S}(x) = \sin (x)$ for a
closed universe, $\mathcal{S}(x) = \sinh (x)$ for an open universe and
$\mathcal{S}(x) = x$ with the factor $\sqrt{|\Omega_k|}$ being removed
for a flat universe.  $H_0$ is the Hubble constant and we sometimes
denote it using a renormalized quantity $h$ defined as $H_0 =100
h~{\rm km ~s}^{-1} {\rm Mpc}^{-1}$. The expansion rate $H(z)$ is given
as
\begin{equation}
H^2(z) = H_0^2 \left[ \Omega_r (1+z)^4 + \Omega_m(1+z)^3 +
  \Omega_k(1+z)^2 + \Omega_X \exp \left( 3 \int_0^z ( 1 +
  w_X(\bar{z})) \frac{d\bar{z}}{1+\bar{z}} \right) \right],
\end{equation}
where $\Omega_r, \Omega_m, \Omega_k$ and
$\Omega_{\rm X}$ represent the present values of the energy density of radiation, matter,
curvature and dark energy normalized by the critical density.  
Note that $\Omega_k = 1 - \Omega_r - \Omega_m - \Omega_X$.
The sound horizon at recombination is
\begin{eqnarray}
r_s(z_{\rm rec}) &=& \int_0^{a_{\rm rec}} \frac{c_s}{a^2 H}\, da
\end{eqnarray}
where $a_{\rm rec} = 1/(1+z_{\rm rec})$ and $c_s$ is the sound speed of the photon-baryon fluid:
\begin{equation}
c_s^2(a) = \frac{\dot{p}_\gamma}{\dot{\rho}_\gamma+\dot{\rho}_b}= \frac{1}{3 (1 + R )}
\end{equation}
with $R = 3 \rho_b /4 \rho_\gamma $ being the scale factor normalized
to $3/4$ at the photon-baryon equality. Since we consider the case
where dark energy is irrelevant at the epoch of recombination, the
above expression can be integrated analytically as
\cite{Hu:1994uz,Hu:2000ti}
\begin{eqnarray}
r_s(z_{\rm rec}) &=& \frac{19.8671}{\sqrt{\omega_b\, \omega_m}} \ln
\left\{ \frac{\sqrt{1+R_{\rm rec}} +\sqrt{R_{\rm rec}+r_{\rm rec}
    R_{\rm rec}} } {1+\sqrt{r_{\rm rec} R_{\rm rec}}} \right\} ~{\rm Mpc},
    \label{eq:sound_horizon}
\end{eqnarray}
where $\omega_b$ and $\omega_m$ are normalized baryon density and matter density
(i.e., $\omega_i = \Omega_i h^2$).  The radiation-to-matter and baryon-to-photon ratios at the epoch of recombination, $r_{\rm rec}$ and $ R_{\rm rec}$,
are given as
\begin{eqnarray}
 r_{\rm rec} &=& 0.042\, \omega_m^{-1}\, (z_{\rm rec}/10^3),\\
R_{\rm rec} &=& 30\, \omega_b \, (z_{\rm rec}/10^3)^{-1}.
\end{eqnarray}
In fact, the redshift at the epoch of recombination slightly
depends on the energy density of baryon and matter. We include the
dependence by adopting the fitting formula $z_{\rm rec}$
\cite{Hu:1995en}:
\begin{eqnarray}
z_{\rm rec} &=& 1048 \left[ 1 + 0.00124\,\omega_b^{-0.738} \right]
\left[ 1 + g_1\, \omega_m^{g_2} \right],
\end{eqnarray}
where the functions $g_1$ and $g_2$ are given as 
\begin{eqnarray}
g_1 &=& 
0.0783\,\omega_b^{-0.238} \left[ 1+ 39.5\, \omega_b^{0.763} \right]^{-1}, \\
g_2 &=& 0.560 \left[ 1+ 21.1\, \omega_b^{1.81} \right]^{-1}.
\end{eqnarray}
We fix the value of $\omega_b$ with the mean value from the WMAP3
analysis $\omega_b = 0.02229$ which is obtained for the $\Lambda$CDM
model.

Once we give the Hubble constant, matter density, baryon density and
dark energy parameters, $\theta_A$ can be calculated.  To constrain
the cosmological parameters, we use the value of $\theta_A$ reported
by WMAP3 for $\Lambda$CDM
\begin{eqnarray}
\theta_{A, {\rm obs}} &=& 0.5952 \,^\circ \pm 0.0021 \,^\circ. \label{eq:theta_A}
\end{eqnarray}
In addition to $\theta_A$, we also use the prior for $\omega_m$ which
is given by WMAP3 for $\Lambda$CDM as
\begin{eqnarray}
\omega_{m, {\rm obs}} &=& 0.1277 \pm 0.008.
\end{eqnarray}
Notice that the values of $\theta_{A, {\rm obs}}$ and $\omega_{m, {\rm obs}}$ are almost unchanged even if
we consider a model with a constant equation of state or a non-flat
universe \cite{Spergel:2006hy,Wang:2007mz}.

Using $\theta_A$ and $\omega_m$, we calculate the $\chi^2$ from CMB as
\begin{eqnarray}
\chi^2_{\rm CMB} = \frac{(\theta_A - \theta_{A, {\rm obs}})^2}{\sigma_{\theta_A}^2} +\frac{(\omega_m - \omega_{m, {\rm obs}})^2}{\sigma_{\omega_m}^2}. 
\label{eq:chi2_CMB}
\end{eqnarray}
Roughly speaking, in this way, although
we are not using the whole information
of the CMB power spectrum but using only the information of the first peak
position and height (after we extracted $\omega_b$ from the second
peak height to calculate $\theta_A$),   since $\theta_A$ and $\omega_m$
are determined from such different, horizontal and vertical, features
of the CMB power spectra, the $\chi^2$ for the parameters we concern
in this paper can be constructed as above. This is explicitly
justified in Sec.~\ref{sec:result} for the flat $\Lambda$CDM
model. Furthermore, since the effects of the curvature of the universe
and dark energy equation of state on the CMB power spectrum mainly
appear as the change of $\theta_A$\footnote{
  The curvature of the universe and dark energy equation of state also 
  affect CMB power spectrum on large scales through the late-time
  integrated Sachs-Wolfe effect, but since the error from the
  cosmic variance dominates on large scales, such an effect can be
  neglected in most cases \cite{Hu:1994jd,Hu:1998kj}. On the other hand, it should be mentioned
  that such large scale fluctuation may be interesting when one
  considers scenarios such as dark energy isocurvature fluctuation
  \cite{Moroi:2003pq,Gordon:2004ez}, the nature of dark energy
  perturbation \cite{Hu:1998kj,Bean:2003fb,Weller:2003hw,Koivisto:2005mm,Ichiki:2007vn} and so on.
}, we can use this $\chi^2_{\rm CMB}$ for constraints involving these
parameters.

\subsection{BAO} \label{sec:method_BAO}

We also use the baryon acoustic oscillation scale measured by SDSS
\cite{Eisenstein:2005su}. To take the BAO data into account, we use
the parameter $D_V$ which is defined as \cite{Eisenstein:2005su}
\begin{equation}
D_V (z) = \left[ r_\theta (z)^2 \frac{z}{H(z)} \right]^{1/3},  \label{eq:DV}
\end{equation}
where $r_\theta(z)$ is the comoving angular diameter distance defined
in the previous subsection\footnote{
  In most of works of dark energy, the so-called $A$ parameter is
  used.  However, we use $D_V(z)$ which explicitly depends on the
  Hubble constant for the purpose of this paper.  The relation between
  $D_V(z)$ and the $A$ parameter is
\begin{equation}
D_V(0.35) 
= 
\frac{A \times 0.35}{\sqrt{\Omega_m H_0^2}} 
= 
3.0 \times 10^3 
\left( 
\frac{A \times 0.35}{\sqrt{\omega_m}} 
\right)
~{\rm Mpc}.  \nonumber
\end{equation}
}.
This quantity depends on the scalar spectral index $n_s$ and
$\omega_b$ slightly. We take $n_s = 0.958$ and $\omega_b = 0.02229$
which are the mean values for the $\Lambda$CDM model from WMAP3 data
alone and calculate $D_V(0.35)$ following the procedure of
Ref.~\cite{Eisenstein:2005su} to find $D_V(0.35) =
1402$\,Mpc. Then we use
\begin{eqnarray}
D_V(0.35)_{\rm obs} = 1402 \pm 64 \, {\rm Mpc},
\end{eqnarray}
for the constraint from BAO.
The value of $\chi^2$ is calculated as 
\begin{eqnarray}
\chi^2_{\rm BAO} = \frac{(D_V(0.35) - D_V(0.35)_{\rm obs})^2}{\sigma_{D_V(0.35)}^2}.
\end{eqnarray}

We note that the measurement of $D_V(0.35)$ assumes a flat $\Lambda$CDM model and there is no explicit check
that the value does not change in a non-flat model and/or dark energy models with $w_X \neq -1$.
However, since the effect of non-flatness or dark energy is considered to be only the geometrical one 
(their effects on the perturbation do not affect the peak of the galaxy correlation function which is located at
smaller scales), we can use $D_V(0.35)$ to probe the background evolution just as we use $\theta_A$ of CMB
if the cosmology is not radically different from a flat $\Lambda$CDM model. 
This condition is considered to be satisfied in our analysis because we place the prior Eq.~\eqref{eq:eos_prior} discussed in the previous section 
and our resulting constraints as shown later are not far away from a flat $\Lambda$CDM model when we combine with CMB and SN.

\subsection{SN}
As for SN data, we calculate the distance modulus
\begin{equation}
\mu = m-M = 5\log d_L + 25,   \label{eq:mu}
\end{equation}
where $m$ is apparent magnitude and $M$ is absolute magnitude.  Here
$d_L$ is the luminosity distance in units of Mpc which is written as
\begin{equation}
d_L(z) = \frac{1+z}{H_0 \sqrt{|\Omega_k|} }
\mathcal{S}  \left( \sqrt{|\Omega_k|} \int_0^z \frac{dz'}{H(z') /H_0} \right).
\end{equation}
Although the luminosity distance depends on the Hubble constant
explicitly, its dependence is indistinguishable with the uncertainty
in $M$.  Thus when we calculate $\chi^2$, the dependence on $h$
vanishes by marginalizing over $M$ as a nuisance parameter.

For the analysis, we use 182 SNe from the Gold data set of Riess et
al. (Gold06) \cite{Riess:2006fw} or 192 SNe of Davis et al. (Davis07)
\cite{Riess:2006fw,Wood-Vasey:2007jb,Davis:2007na}.  In this paper, we
present our results for these two data sets separately.  (Namely we do
not combine these two data sets.)

The main difference between these two data sets is that Davis07 uses the first data release of the ESSENCE supernova survey
\cite{Wood-Vasey:2007jb,Miknaitis:2007jd} at the intermediate redshift. 
In detail, Gold06 consists of 119 SNe (38 SNe are nearby, $z \lesssim 0.05$) from the previous Gold data set \cite{Riess:2004nr}, 16 SNe which are recently discovered by Hubble Space Telescope (HST) \cite{Riess:2006fw} and 47 SNe from the first data release of the SNLS project \cite{Astier:2005qq}.
Davis07 includes 60 SNe from ESSENCE, 57 SNe from SNLS, 16 SNe of Ref.~\cite{Riess:2006fw}, 14 SNe discovered by HST
which are included in the previous Gold data set and 45 nearby ($z \lesssim 0.05$) SNe. Thus, 30 SNe which are discovered by HST
and have relatively high redshifts are common in both data sets. Also there are 25 nearby SNe which they have in common.
The SNLS first data release \cite{Astier:2005qq} has 73 SNe but Gold06 and Davis07 place a different criterion for the selection.
We may roughly say that Davis07 is constructed from Gold06 by replacing SNe data at the intermediate redshifts which are discovered in earlier times by
the recent ESSENCE data.

\subsection{Hubble constant} \label{subsec:method_H}
One of the purposes of this paper is to constrain the Hubble constant
using the cosmological observations introduced above. However, as
mentioned in the introduction, the Hubble constant can be measured
with other methods such as using the cosmic distance ladder
measurements. We summarize the recent values obtained with the
distance ladder in Table~\ref{tab:summary_h}. 
For a recent review on the Hubble constant, see e.g. Ref.~\cite{Jackson:2007ug}
\footnote{
In particular, interesting constraints on the Hubble constant are obtained via 
gravitational lens time delays  and the Sunyaev-Zel'dovich effect,
although the uncertainties are comparable or slightly larger than those of the distance ladder measurements.
See Refs.~\cite{Oguri:2006qp} and \cite{Bonamente:2005ct} for the recent measurements by these methods. 
Since these two methods in principle depend also on a dark energy model, their future improvements 
are expected to be useful for probing the cosmological constraints on the Hubble constant and dark energy 
parameters independently from CMB, BAO and SN. 
}. 
We would like to compare
such $H_0$ values measured in a relatively direct way with those
obtained from the measurements of CMB, BAO and SN. Furthermore we
also investigate constraints on dark energy putting a prior on the
Hubble constant to see how the determination of the Hubble parameter
affects the constraints on dark energy. Thus we also use these values
as priors when deriving constraints on dark energy parameters and/or
the curvature of the universe. When we use the prior on the Hubble
constant, we include them by calculating the $\chi^2$ which is given
by
\begin{eqnarray}
\chi^2_{H_0} = \frac{(H_0 - H_{0,{\rm obs}})^2}{\sigma_{H_0}^2}.
\end{eqnarray}
For the error $\sigma_{H_0}$, since we do not obtain a significant
effect with those of the current measurements given in Table
\ref{tab:summary_h}, as will be briefly discussed in the following
section (Sec.~\ref{subsec:constraint_DE}), we take a hypothetical
value $\sigma_{H_0} = 2$\,km\,s$^{-1}$\,Mpc$^{-1}$.  
This value is motivated by the expected accuracy of 1\% \cite{Macri:2006wm} which could be obtained
through measuring maser distance to a large number of galaxies in the
Hubble flow by planned radio telescopes such as Square Kilometer Array (SKA) \cite{SKA}. 
We adopt a slightly more conservative value than that.

  \begin{table}
  \centering 
  \begin{tabular}{|l|c|}
  \hline
Refs.  & $H_0$ $\pm$(statistic) $\pm$ (systematic)  \\
  \hline
 Freedman et al. \cite{Freedman:2000cf} & $72 \pm 3 \pm 7$ \\
 Sandage et al. \cite{Sandage:2006cv} & $62.3 \pm 1.3 \pm 5.0$ \\
 Macri et al. \cite{Macri:2006wm} & $74 \pm 3 \pm 6$ \\
   \hline
  \end{tabular}
  \caption{The values of the Hubble constant in unit of 
    ${\rm km\,s}^{-1} {\rm Mpc}^{-1}$ determined from cosmic distance 
    ladder measurements.}
    \label{tab:summary_h}
  \end{table}

\newpage 

\section{Results} \label{sec:result} 

Now we discuss the constraints on the Hubble constant assuming some
dark energy models including time-evolving dark energy equations of state described in the
previous section. 
We also investigate how the constraint on dark
energy parameters are affected by the assumption of the Hubble
constant. Effects of the curvature of the universe in constraining
these parameters and the constraint on itself are also investigated.

\subsection{Constraint on the Hubble constant}  \label{subsec:constraint_H}

\subsubsection{Case with a cosmological constant} \label{subsubsec:cc}

\begin{figure}
    \centerline{\scalebox{1}{\includegraphics{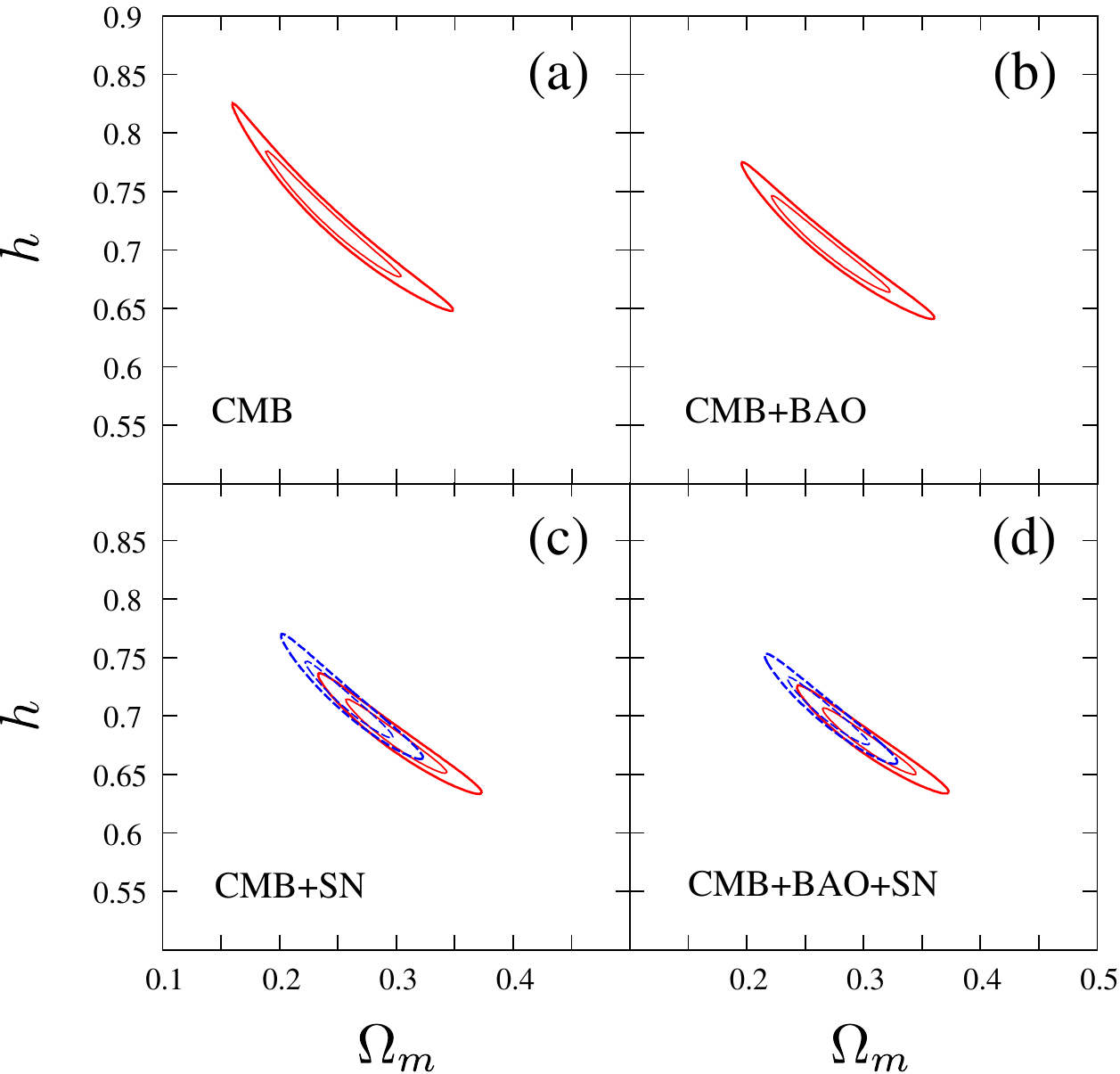}}}
    \caption{Contours of 1$\sigma$ and 2$\sigma$ allowed regions in
      the $\Omega_m$--$h$ plane for the case with a cosmological
      constant in a flat universe.  (a) CMB alone, (b) CMB+BAO, (c)
      CMB+SN and (d) CMB+BAO+SN.  In the panels (c) and (d), we treat
      the SN data from Gold06 (red solid line) and Davis07 (blue
      dashed line) separately, thus two different constraints are
      shown.}
\label{fig:Om_h_cc}
\end{figure}

First we show the constraints in the $\Omega_m$--$h$ plane for the case with a cosmological
constant in Fig.~\ref{fig:Om_h_cc}.  In the figure, we show the
constraints from several data sets, i.e., the cases with CMB alone,
CMB+BAO, CMB+SN and CMB+BAO+SN.  We do not show the constraint from SN
data alone and BAO data alone here.  Since, for SN data, the
dependence of the Hubble constant on the luminosity distance is
absorbed into the uncertainty in the absolute magnitude of SN, which
is marginalized over, we cannot obtain the information on $H_0$ from
SN alone.  For the constraint from BAO, we use the parameter $D_V$ for
the analysis in this paper, which cannot constrain the Hubble constant
much by itself from the currently available data. Thus we only
show the constraints from CMB alone and the combinations of CMB and
other data sets.

For the analysis with CMB data alone, our method gives consistent
results with those obtained by the WMAP team who uses information of
the entire CMB power spectrum.  For the marginalized 1$\sigma$ values
of $h$ and $\Omega_m$, theirs are $h=0.732^{+0.031}_{-0.032}$ and
$\Omega_m = 0.241 \pm 0.034$, which are satisfactorily close to our
values $h=0.729 \pm 0.035$ and $\Omega_m = 0.244 \pm 0.038$.  Closer
comparison reveals that their two-dimensional contours in this plane
(Fig.~10 in Ref.~\cite{Spergel:2006hy}) are somewhat larger than ours
(Fig.~\ref{fig:Om_h_cc} (a)) but the difference scarcely affects the
estimation of $h$ and $\Omega_m$ as mentioned above. 
This small difference arises most likely because we fix $\omega_b$ to
the WMAP central value and our method cannot incorporate the
uncertainty due to the value of $n_s$.  As for the SN data sets, we show
our results with Gold06 and Davis07 separately.  In Table
\ref{tab:h_cc}, we report the central values and 1$\sigma$ errors from
various combinations of data sets for the case with a cosmological
constant.  The values of the minimum $\chi^2$ for each case are also
shown.  When we combine SN with CMB, the central value of $h$ becomes
smaller since the constraint on $\Omega_m$ and $h$ lies along the
region around $\Omega_m h^2$ being constant (for the case with WMAP3,
$\Omega_m h^2= 0.1277$) and SN data favor somewhat larger $\Omega_m$
than CMB.  This is more conspicuous for Gold06 than Davis07.  Another
point is that the BAO data also favors a slightly smaller value of
$h$.  Hence when we use the data set of CMB+BAO+SN(Gold06), the
favored value of $h$ becomes considerably smaller.

\begin{table}
  \centering 
  \begin{tabular}{|l||c|c|c|c|}
  \hline
  & $h$ ($1\sigma$)  & $\Omega_m$ ($1\sigma$) & $\chi^2_{\rm min}$ & \# of data \\
  \hline
  CMB alone  & $0.729 \pm 0.035$ & $0.244 \pm 0.038$ & 0.00  & 2 \\
  CMB+BAO  & $0.704 \pm 0.027$ & $0.270 \pm 0.033$ & 1.41   & 3 \\
  CMB+SN(Gold06)  & $0.682 \pm 0.021$ & $0.299 \pm 0.028$ & 162.0  &  184 \\
  CMB+SN(Davis07)  & $0.714 \pm 0.022$ & $0.259 \pm 0.025$& 195.9  &  194 \\
  CMB+BAO+SN(Gold06)  & $0.678 \pm 0.019$ & $0.304 \pm 0.026$&162.2 & 185 \\
 CMB+BAO+SN(Davis07) & $0.704 \pm 0.019$ & $0.269 \pm 0.023$ &197.0 & 195  \\
   \hline
  \end{tabular}
  \caption{The central values  and 1$\sigma$ errors 
    from several combinations of data sets for the case with a
    cosmological constant.  The values of the minimum $\chi^2$ and the numbers of data are also shown.}
  \label{tab:h_cc}
\end{table}

\subsubsection{Case with a constant equation of state} \label{subsubsec:const_w}
\begin{figure}
    \centerline{\scalebox{1}{\includegraphics{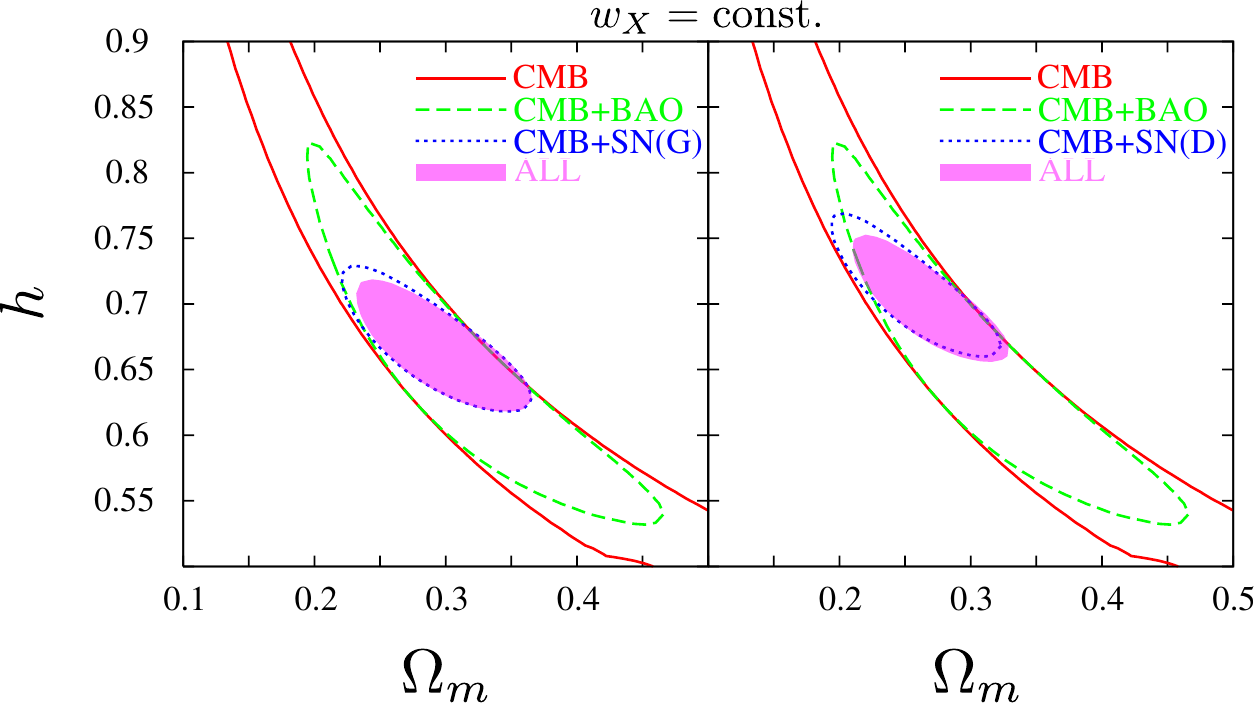}}}
    \caption{2$\sigma$ constraints are shown for the case with a
      constant equation of state for dark energy.  We use the data
      from CMB alone (red solid line), CMB+BAO (green dashed line),
      CMB+SN (blue dotted line) and all combined (purple shaded
      region). As for the treatment of SN data, we use those from
      Gold06 (left panel) and Davis07 (right panel).  The value of
      $w_0$ is marginalized over $-3 \le w_0 \le -1/3$.}
      \label{fig:Om_h_margw0}
\end{figure}

Next we show the constraints for the case with a constant equation of
state for dark energy (which corresponds to the cases with $w_1=0$ in
Eq.~(\ref{eq:eos}) and with $\tilde{w}_0 = \tilde{w}_1$ in
Eq.~\eqref{eq:eos2}).  In Fig.~\ref{fig:Om_h_margw0}, the constraints
in the $\Omega_m$--$h$ plane are shown after marginalizing over
$w_0$. In the figure, 2$\sigma$ constraints from CMB alone (red solid
line), CMB+BAO (green dashed line), CMB+SN (blue dotted line) and all
combined (purple shaded region) are shown.  We treat the SN data from
Gold06 and Davis07 separately, thus we show the constraints using them
in the left and right panels in Fig.~\ref{fig:Om_h_margw0} respectively.

Since dark energy equation of state directly affects the angular
diameter distance, the allowed region from CMB alone becomes
significantly larger compared to the case with a cosmological
constant, which means that there is a strong degeneracy among
$\Omega_m$, $w_0$ and $h$. However, if we combine other data set such
as BAO and SN with CMB data, the allowed region becomes similar to
that for the case with a cosmological constant (the region becomes
larger but only a little).  Although the preferred values of $w_0$
which is marginalized over in the figure are in general different for
each observation for fixed $h$ and $\Omega_m$, only around the allowed
region in the figure, those values from
different observations happen to coincide to be $w_0 \sim-1$ (i.e. a cosmological constant).  
Thus we obtain an almost same result as the cosmological constant case even for
the case where we marginalize over a constant equation of state.  We see that the
combination of the data from CMB+SN seems to be enough to constrain
$h$ and $\Omega_m$ whereas the CMB+BAO constraint is much weaker.
This is because the degeneracy curves in the $\Omega_m$--$w_0$ plane
extend to almost the same direction for CMB and BAO, but those of CMB
and SN are complementary (see Fig.~6 in
Ref.~\cite{Ichikawa:2005nb})\footnote{
  In Ref.~\cite{Ichikawa:2005nb}, the so-called $A$ parameter has been used
  to obtain a constraint from BAO, whereas $D_V$ is used in this
  paper.  However, as long as we also use the prior on $\omega_m$,
  these two constraints are almost equivalent.
  Also, the shift parameter $R$ has been used to obtain a constraint from CMB
  but this gives a similar result to our method of combining $\theta_A$ and $\omega_m$.
}.  Thus as far as the degeneracy between $\Omega_m$ and $w_0$ is
concerned, the combination of CMB+SN gives a severe constraint.

\subsubsection{Case with time-evolving equations of state}  \label{subsubsec:evolving_w}
\begin{figure}
    \centerline{\scalebox{1}{\includegraphics{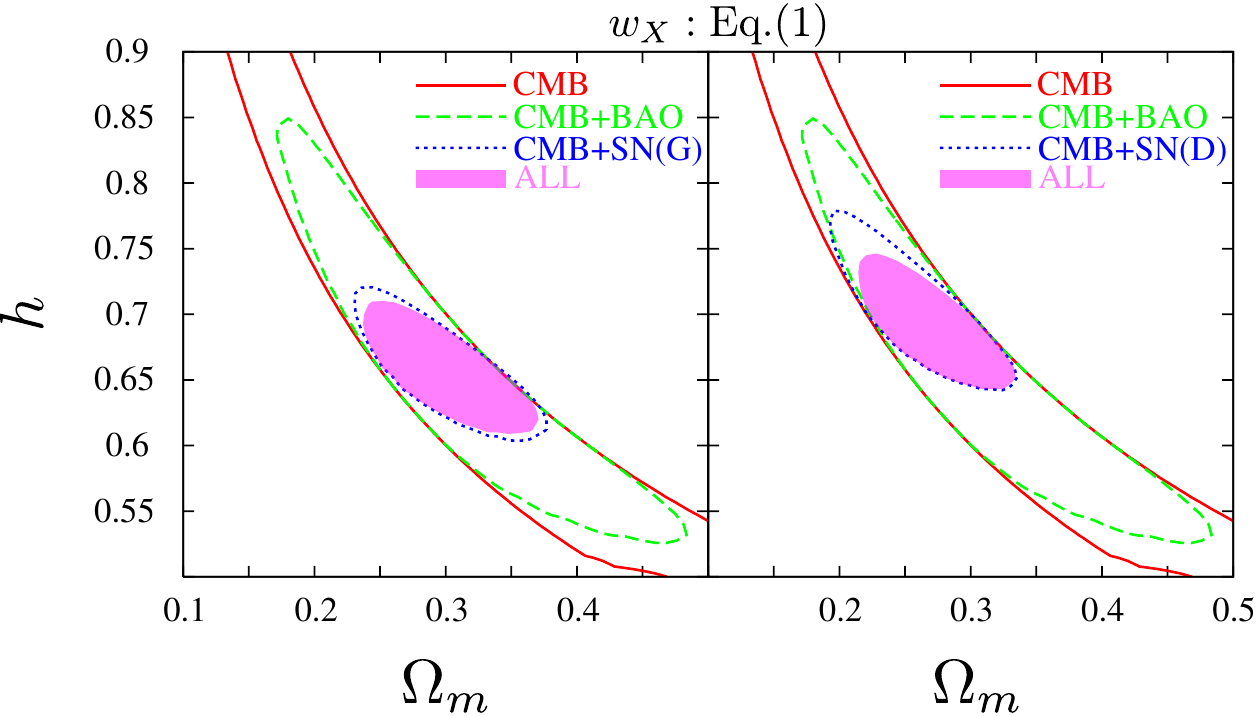}}}
    \caption{Same as Fig.~\ref{fig:Om_h_margw0} except we assume
      the dark energy equation of state as Eq.~\eqref{eq:eos} and
      marginalize over the values of $w_0$ and $w_1$ in the ranges $-3 \le w_0 \le
      -1/3$ and $-3 \le w_1 \le 3$ with the prior
      Eq.~\eqref{eq:eos_prior2}. }
      \label{fig:Om_h_margw0w1}
\end{figure}

In Fig.~\ref{fig:Om_h_margw0w1}, we show the results for the case with
the time-evolving dark energy equation of state parametrized as Eq.~\eqref{eq:eos}. 
In the figure, the values of $w_0$ and $w_1$ are marginalized over. As the
case with a constant equation of state, although the allowed region
from CMB data alone is significantly larger compared to that for a
cosmological constant case, when other data are combined, the allowed
region lies around the same region as the $\Lambda$CDM case and also
it is not so significantly larger compared to that for the
$\Lambda$CDM model. Notice that the combination of CMB+SN again
already gives severe constraint in the $\Omega_m$--$h$ plane.  This
result shows that even for a time-evloving equation of state, the SN data
works well to break the degeneracy among $\Omega_m$ and dark energy
parameters for the parametrization Eq.~\eqref{eq:eos}.

\begin{figure}
    \centerline{\scalebox{1}{\includegraphics{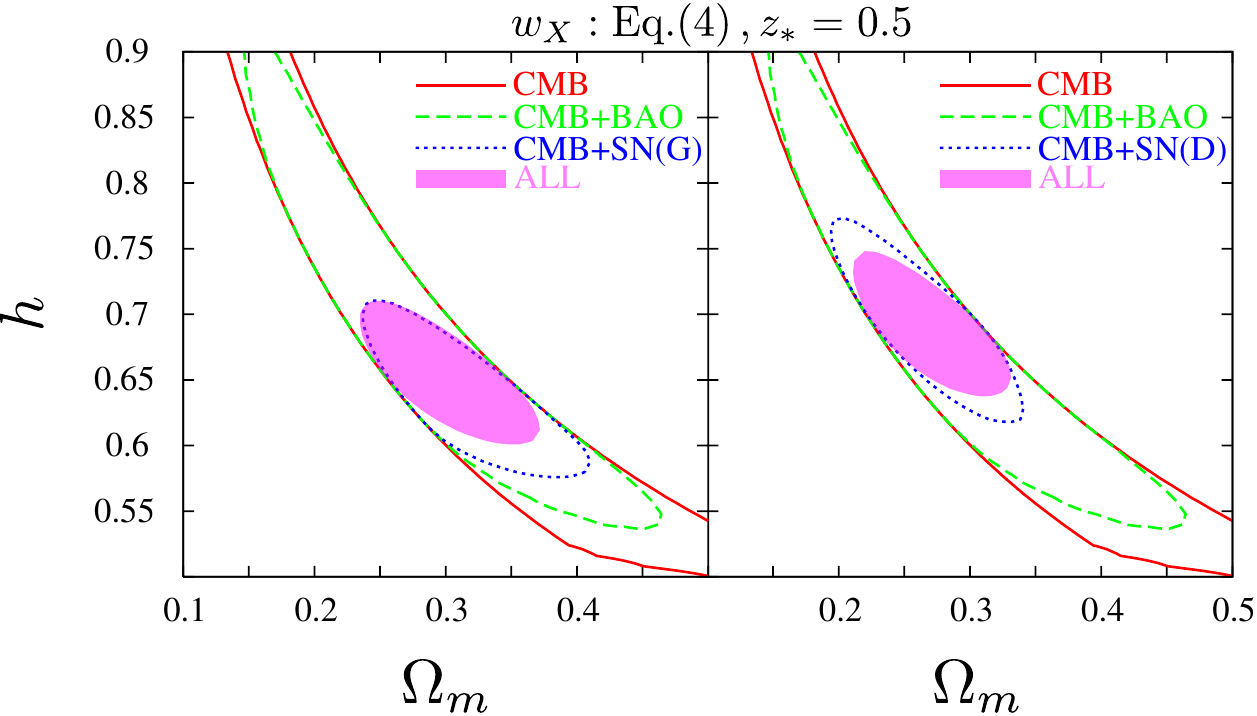}}}
    \caption{Same as Fig.~\ref{fig:Om_h_margw0} except we assume
      the dark energy equation of state as Eq.~\eqref{eq:eos2} with
      $z_\ast=0.5$ and marginalize over the values of $\tilde{w}_0$
      and $\tilde{w}_1$ in the ranges $-3 \le \tilde{w}_0 \le -1/3$ and $-3 \le
      \tilde{w}_1 \le -1/3$. }
      \label{fig:Om_h_margw0w1zast}
\end{figure}

We also performed the same analysis using the parametrization given in
Eq.~\eqref{eq:eos2} with $z_\ast = 0.5$ and the results are shown in
Fig.~\ref{fig:Om_h_margw0w1zast}.  As seen from the figure, the
constraints on $\Omega_m$ and $h$ are almost the same even if we assume
this type of dark energy parametrization.  Furthermore it should be
noticed that here again the combination of CMB+SN gives a severe
constraint.

Although we have adopted just two types of parametrization of dark
energy equation of state, as far as we consider some typical types of
dark energy evolution, it seems that dark energy models do not affect
much the determination of the Hubble constant using combined data of
CMB, BAO and SN. As already mentioned and seen from
Figs.~\ref{fig:Om_h_margw0}, \ref{fig:Om_h_margw0w1} and
\ref{fig:Om_h_margw0w1zast}, CMB+SN gives sufficiently tight
constraints in the $\Omega_m$--$h$ plane. Meanwhile, the allowed
region from CMB+BAO is rather broad and BAO does not seem to have much
constraining power.  (However, when we allow a non-flat universe, BAO
becomes important, which we are going to discuss in
Sec.~\ref{subsubsec:nonflat}.) Thus we focus on the constraints from
CMB and SN for a while and explain how the constraints can be
obtained.

\begin{table}
  \centering
  \begin{tabular}{|l||c|c|c|c|c|}
  \hline
CMB+BAO+Gold06 & $h$ ($1\sigma$) & $\Omega_m$ ($1\sigma$)  &$\chi^2_{\rm min}$ & $w_0$ ($\tilde{w}_0$) & $w_1$ ($\tilde{w}_1$) \\
  \hline
  Cosmological constant  & $0.678 \pm 0.019$ & $0.304 \pm 0.026$ & 162.2  & $-$ & $-$ \\
  $w_0$ ($w_1=0$) & $0.667 \pm 0.020 $ & $0.295 \pm 0.027$ &160.2  & -0.87 & $-$ \\
  $w_0$ and $w_1$  & $0.656 \pm 0.020$ & $0.301 \pm 0.026$ &158.3  & -1.06 & 0.72\\
  $w_0$ and $w_1$ ($w \ge -1$) & $0.656 \pm 0.020$ & $0.299 \pm 0.027$ &  158.5 & -1.00 & 0.60 \\
  $\tilde{w}_0$ and $\tilde{w}_1$ ($z_\ast = 0.1$) &$0.678 \pm 0.023$ & $0.279 \pm 0.027$ & 157.9  & -2.49 & -0.71 \\
  $\tilde{w}_0$ and $\tilde{w}_1$ ($z_\ast = 0.2$) &$0.666 \pm 0.021$ & $0.289 \pm 0.026$ & 157.5 & -1.63 & -0.68 \\
  $\tilde{w}_0$ and $\tilde{w}_1$ ($z_\ast = 0.5$) &$0.655 \pm 0.022$ & $0.299 \pm 0.027$ & 157.5 & -1.21 & -0.61 \\
  $\tilde{w}_0$ and $\tilde{w}_1$ ($z_\ast = 1.0$) &$0.649 \pm 0.023$ & $0.306 \pm 0.029$ & 158.1 & -1.09 & -0.52 \\
  $\tilde{w}_0$ and $\tilde{w}_1$ ($z_\ast = 2.0$) &$0.651 \pm 0.022$ & $0.307 \pm 0.029$ & 158.6 & -1.02 & -0.37 \\
  $\tilde{w}_0$ and $\tilde{w}_1$ ($z_\ast = 0.5$) ($w \ge -1$) &$0.658 \pm 0.022$ & $0.294 \pm 0.027$ & 158.3 & -1.00 & -0.71 \\
  $\tilde{w}_0$ and $\tilde{w}_1$ ($z_\ast = 1.0$) ($w \ge -1$) &$0.652 \pm 0.023$ & $0.301 \pm 0.027$ & 158.3 & -1.00 & -0.60 \\
  $\tilde{w}_0$ and $\tilde{w}_1$ ($z_\ast = 2.0$) ($w \ge -1$) &$0.651 \pm 0.023$ & $0.305 \pm 0.028$ & 158.6 & -1.00 & -0.41 \\
   \hline
   \hline
CMB+BAO+Davis07 & $h$ ($1\sigma$) & $\Omega_m$ ($1\sigma$)  &$\chi^2_{\rm min}$ & $w_0$ ($\tilde{w}_0$)  & $w_1$ ($\tilde{w}_1$)  \\
  \hline
  Cosmological constant  & $0.704 \pm 0.019$ & $0.269 \pm 0.023$ & 197.0  & $-$ & $-$ \\
  $w_0$ ($w_1=0$) & $0.703 \pm 0.020 $ & $0.266 \pm 0.024$ & 196.9  &-0.98 & $-$ \\
  $w_0$ and $w_1$  & $0.689 \pm 0.025$ & $0.272 \pm 0.024$ & 195.5  & -1.16 & 0.83\\
  $w_0$ and $w_1$ ($w \ge -1$) & $0.695 \pm 0.022 $ & $0.265 \pm 0.024 $ & 196.4 & -1.00 & 0.27 \\
  $\tilde{w}_0$ and $\tilde{w}_1$ ($z_\ast = 0.1$) &$0.707 \pm 0.021$ & $0.259 \pm 0.024$ & 195.5 & -2.03 & -0.85 \\
  $\tilde{w}_0$ and $\tilde{w}_1$ ($z_\ast = 0.2$) &$0.700 \pm 0.020$ & $0.264 \pm 0.024$ & 195.9 & -1.40 & -0.84 \\
  $\tilde{w}_0$ and $\tilde{w}_1$ ($z_\ast = 0.5$) &$0.693 \pm 0.023$ & $0.268 \pm 0.024$ & 195.9 & -1.18 & -0.77 \\
  $\tilde{w}_0$ and $\tilde{w}_1$ ($z_\ast = 1.0$) &$0.684 \pm 0.025$ & $0.274 \pm 0.025$ & 195.5 & -1.15 & -0.60 \\
  $\tilde{w}_0$ and $\tilde{w}_1$ ($z_\ast = 2.0$) &$0.682 \pm 0.024$ & $0.277 \pm 0.026$ & 195.4 & -1.12 & -0.38 \\
  $\tilde{w}_0$ and $\tilde{w}_1$ ($z_\ast = 0.5$) ($w \ge -1$) &$0.698 \pm 0.020$ & $0.264 \pm 0.024$ & 196.6 & -1.00 & -0.90 \\
  $\tilde{w}_0$ and $\tilde{w}_1$ ($z_\ast = 1.0$) ($w \ge -1$) &$0.693 \pm 0.023$ & $0.265 \pm 0.023$ & 196.3 & -1.00 & -0.80 \\
  $\tilde{w}_0$ and $\tilde{w}_1$ ($z_\ast = 2.0$) ($w \ge -1$) &$0.689 \pm 0.025$ & $0.267 \pm 0.023$ & 196.2 & -1.00 & -0.63 \\
   \hline
  \end{tabular}
  \caption{The central values and 1$\sigma$ errors, along with 
    the minimum value of $\chi^2$  from CMB+BAO+SN 
    for the cases with various dark energy models. 
    Here the dark energy parameters 
    are marginalized over. 
    The cases where  the equation of state is always larger than $-1$ are 
    also presented.
    A flat universe is assumed. 
    The best-fit values of dark energy parameters are also shown.
  }
\label{tab:margDE}
\end{table}

Notice that since $h$ and $\Omega_m$ are degenerate with respect to
$\theta_A$, to break the degeneracy, we assume the prior on $\omega_m$
determined from the height of the CMB power spectrum. Thus the fact
that $h$ is well determined even if we allow some variations of dark
energy models eventually indicates that the determination of
$\Omega_m$ is not affected much by the assumption for dark energy.
(Also notice that, as already mentioned, the SN data cannot determine the
Hubble constant since the dependence on $h$ in the luminosity distance
is totally indistinguishable from the uncertainty in the absolute magnitude of SN.)  Remind that
the SN data can give information of the background evolution from the
present up to $z_{\rm SN}$ which represents the redshift of the
furthest SN currently available (in observations used in the anlysis,
$z_{\rm SN} \approx 1.8$). In this period, the background evolution is
determined by dark energy and matter.  Since the distance measure
such as the luminosity distance
involves some integration with respect to redshift, it is well-known
that there is a severe degeneracy among $\Omega_m$ and dark energy
parameters.  However we can break the degeneracy by combining the
constraint from $\theta_A$ which can also probe the background
evolution earlier than $z_{\rm SN}$.  In the epoch earlier than
$z_{\rm SN}$, we can approximate that the universe is dominated by
matter component.  Thus, knowing the distance between the present
epoch and $z_{\rm SN}$ by the SN data, the matter density is
determined accurately from the distance measurement between $z_{\rm
  SN}$ and $z_{\rm rec}$.  The point is that the distance between
$z=0$ and $z_{\rm SN}$, where both of the matter and dark
energy contribute, can be inferred by the SN data independently from
dark energy model. This is why the combination of SN and CMB can
determine $\Omega_m$ in a dark energy model-independent manner.  We
will provide a more quantitative demonstration of this point in
Appendix \ref{sec:OmOk_SNCMB}.

As we have just argued, the Hubble constant derived from CMB, BAO and SN is
not affected much by the assumption for a dark energy model.  Notice
that, however, when we impose a prior on the Hubble constant by using
a value for instance of the cosmic distance ladder measurement
mentioned in Sec.~\ref{subsec:method_H}, the constraints on dark
energy parameters can be affected by the prior on $H_0$.  Naturally,
its effect depends on the accuracy of the determination of the Hubble
constant. We will return this issue in
Sec.~\ref{subsec:constraint_DE}.

We summarize our results on the constraints on $h$ and $\Omega_m$ for
time-evolving dark energy equations of state in Table~\ref{tab:margDE}
along with the best-fit values of the dark energy parameters.  As
discussed above, even if we assume a different dark energy
parametrization, the constraints on $h$ and $\Omega_m$ are almost
unchanged.  Furthermore, even if we restrict ourselves to the dark
energy model with its equation of state being larger than $-1$ in the
course of the history of the universe, the results are not affected by
this assumption as seen from Table~\ref{tab:margDE}.  It should be
mentioned that two SN data sets (i.e., Gold06 and Davis07) give
somewhat different values of the Hubble constant as discussed above.
The difference due to the SN data set is more significant than the one
caused by the dark energy parametrization.

\subsubsection{Case with a non-flat universe}  \label{subsubsec:nonflat}
Here we discuss the constraint on the Hubble constant without assuming
a flat universe. When one considers the constraint on the Hubble
constant from cosmological observations, a flat universe is often assumed.  
However, to make our analysis general, here we allow a non-flat universe to obtain the
constraint on the Hubble constant. The constraint on the curvature of
the universe itself with some priors on the Hubble constant will be
discussed in Sec.~\ref{subsec:constraint_Ok}.

 \begin{figure}
    \centerline{\scalebox{1}{\includegraphics{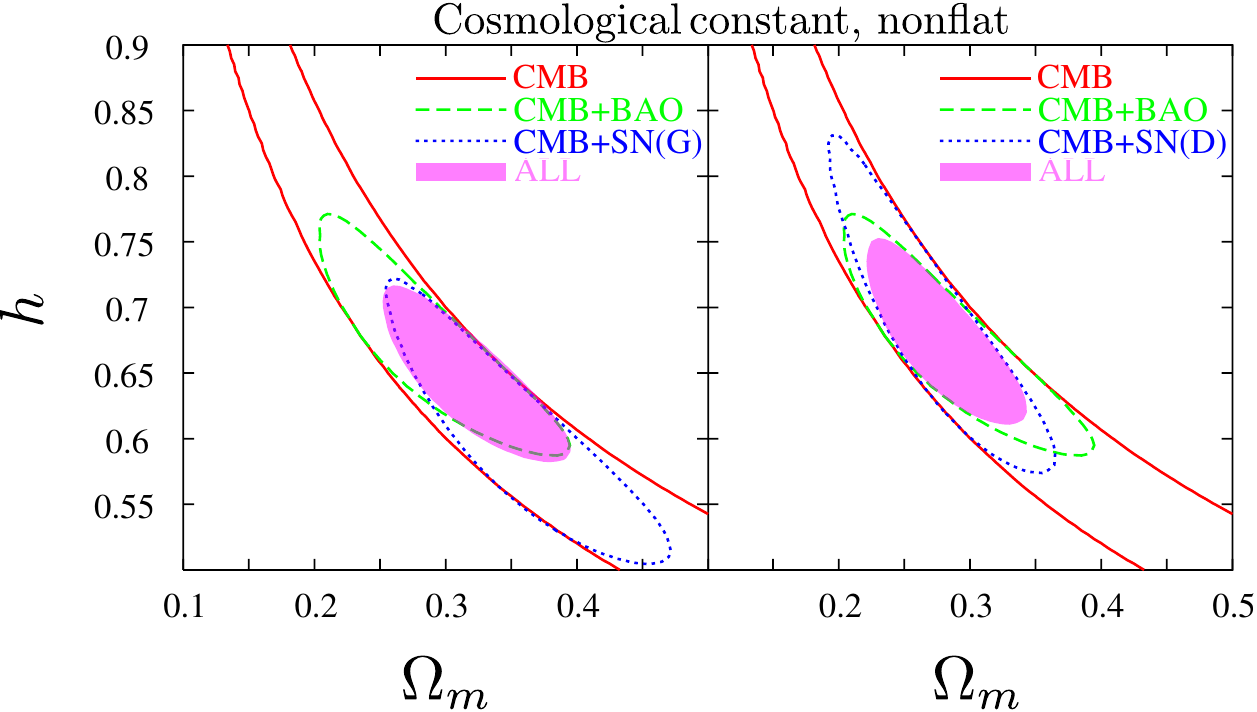}}}
    \caption{2$\sigma$ constraints in the $\Omega_m$--$ h$ plane without
      assuming a flat universe. The curvature is marginalized over in
      the range $-0.5 \le \Omega_k \le 0.5$.  A cosmological constant
      is assumed for dark energy. Constraints from the data from CMB
      alone (red solid line), CMB+BAO (green dashed line), CMB+SN
      (blue dotted line) and all combined (purple shaded region) are
      shown.  As for the treatment of the SN data, we use those from
      Gold06 (left panel) and Davis07 (right panel).  }
	\label{fig:Om_h_margcurv_lambda}
\end{figure}

 \begin{figure}
    \centerline{\scalebox{1}{\includegraphics{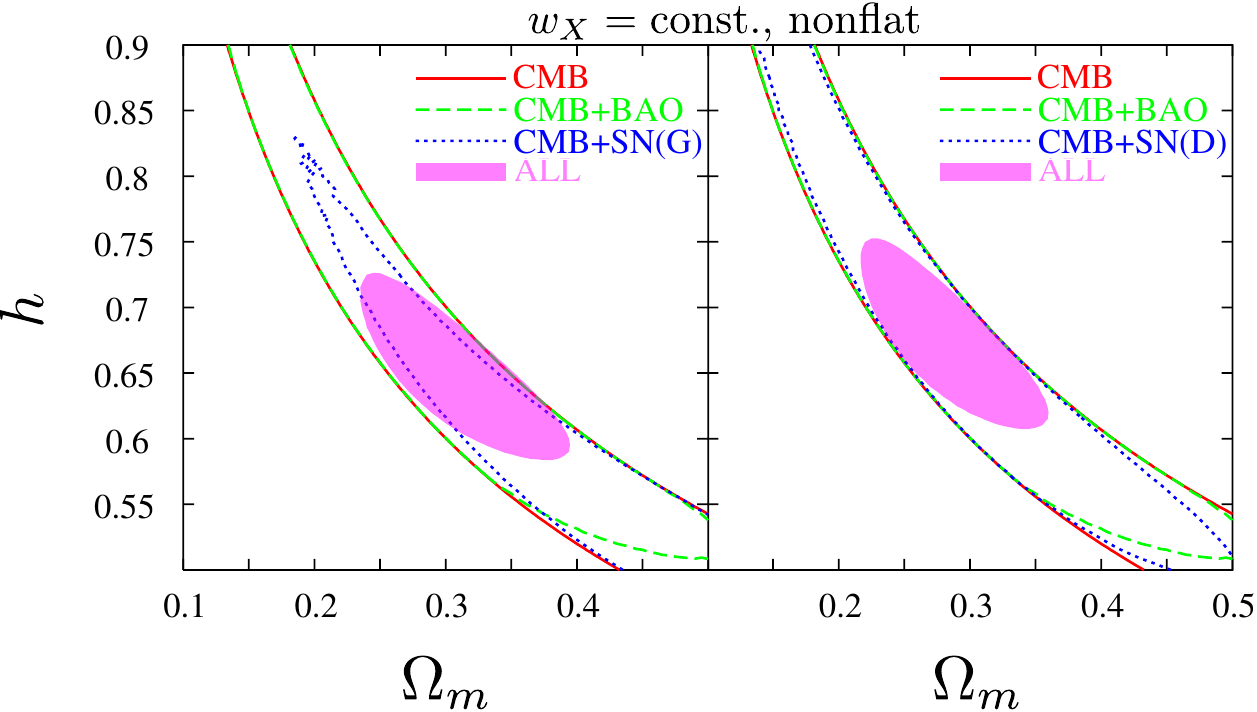}}}
    \caption{Same as Fig.~\ref{fig:Om_h_margcurv_lambda} (i.e. the
      curvature is marginalized over) except that a constant equation
      of state is assumed and marginalized over in the range $-3
      \le w_0 \le -1/3$.}
	\label{fig:Om_h_margcurv_w0}
\end{figure}
     
\begin{table}[htb]
  \centering
  \begin{tabular}{|l||c|c|c|c|c|c|}
  \hline
CMB+BAO+Gold06 & $h$ ($1\sigma$) & $\Omega_m$ ($1\sigma$)  &$\chi^2_{\rm min}$ & $\Omega_k$ & $w_0$ ($\tilde{w}_0$) & $w_1$ ($\tilde{w}_1
$) \\
  \hline
  Cosmological constant  & $0.644 \pm 0.027$ & $0.319 \pm 0.029$ & 159.8  & -0.014 & $-$ & $-$ \\
  $w_0$ ($w_1=0$) & $0.649 \pm 0.029$ & $0.310 \pm 0.032$ & 159.5 & -0.010 &  -0.93 & $-$ \\
  $w_0$ and $w_1$  & $0.656 \pm 0.029$ & $0.301 \pm 0.031$ & 158.3 & -0.000 &  -1.05 & 0.72 \\
  $\tilde{w}_0$ and $\tilde{w}_1$ ($z_\ast = 0.1$) &$0.685 \pm 0.041$ & $0.276 \pm 0.037$ & 157.8  & 0.003 &  -2.62 & -0.69\\
  $\tilde{w}_0$ and $\tilde{w}_1$ ($z_\ast = 0.2$) &$0.680 \pm 0.037$ & $0.279 \pm 0.034$ & 157.3 & 0.009 & -1.80 & -0.59 \\
  $\tilde{w}_0$ and $\tilde{w}_1$ ($z_\ast = 0.5$) &$0.676 \pm 0.032$ & $0.282 \pm 0.030$ & 156.6 & 0.032 &  -1.39 & -0.33 \\
  $\tilde{w}_0$ and $\tilde{w}_1$ ($z_\ast = 1.0$) &$0.662 \pm 0.029$ & $0.295 \pm 0.031$ & 157.5 & 0.018 &  -1.12 & -0.33 \\
  $\tilde{w}_0$ and $\tilde{w}_1$ ($z_\ast = 2.0$) &$0.654 \pm 0.029$ & $0.303 \pm 0.032$ & 158.5 & 0.006 &  -1.02 & -0.33\\
   \hline
   \hline
CMB+BAO+Davis07 & $h$ ($1\sigma$) & $\Omega_m$ ($1\sigma$)  &$\chi^2_{\rm min}$ & $\Omega_k$  & $w_0$ ($\tilde{w}_0$)  & $w_1$ ($\tilde{w}
_1$)  \\
  \hline
  Cosmological constant  & $0.677 \pm 0.028$ & $0.278 \pm 0.025$ & 195.6  & -0.011 &  $-$ & $-$ \\
  $w_0$ ($w_1=0$) & $0.674 \pm 0.029$ & $0.284 \pm 0.029$ & 195.5 & -0.012  &-1.04 & $-$ \\
  $w_0$ and $w_1$  & $0.677 \pm 0.034$ & $0.281 \pm 0.033$ & 195.4 & -0.008  & -1.10 & 0.41 \\
  $\tilde{w}_0$ and $\tilde{w}_1$ ($z_\ast = 0.1$) &$0.688 \pm 0.036 $ & $0.272 \pm 0.033$ & 195.1 & -0.008 &  -1.69 & -0.93\\
  $\tilde{w}_0$ and $\tilde{w}_1$ ($z_\ast = 0.2$) &$0.680 \pm 0.034$ & $0.279 \pm 0.033$ & 195.4 & -0.010 &  -1.20 & -0.97 \\
  $\tilde{w}_0$ and $\tilde{w}_1$ ($z_\ast = 0.5$) &$0.676 \pm 0.034$ & $0.282 \pm 0.033$ & 195.5 & -0.011 &  -1.07 & -0.98 \\
  $\tilde{w}_0$ and $\tilde{w}_1$ ($z_\ast = 1.0$) &$0.678 \pm 0.034$ & $0.281 \pm 0.033$ & 195.4 & -0.008  &  -1.09 & -0.83 \\
  $\tilde{w}_0$ and $\tilde{w}_1$ ($z_\ast = 2.0$) &$0.677 \pm 0.032$ & $0.282 \pm 0.032$ & 195.4 & -0.005  &  -1.10 & -0.53\\
   \hline
  \end{tabular}
  \caption{The central values and 1$\sigma$ errors, along with 
    the minimum value of $\chi^2$  from CMB+BAO+SN 
    for the cases with various dark energy models. Here the dark energy parameters 
    and $\Omega_k$ are marginalized over.
    The best-fit dark energy parameters and $\Omega_k$ are also shown.
  }
  \label{tab:margDE_Ok}
  \end{table}

In Fig.~\ref{fig:Om_h_margcurv_lambda}, we show the constraints in the
$\Omega_m$--$h$ plane for the case with a cosmological constant while
marginalizing over $\Omega_k$ in the range $-0.5 \le \Omega_k \le
0.5$.  We show the constraints from CMB alone (red solid line),
CMB+BAO (green dashed line), CMB+SN (blue dotted line) and all
combined (purple shaded region) separately.  The case with
marginalizing over a constant equation of state in addition is
presented in Fig.~\ref{fig:Om_h_margcurv_w0}.  When we allow a
non-flat universe, there is a strong degeneracy in the distance measures
as in the case we assume a dynamical dark energy in a flat universe
which can be seen in Figs.~\ref{fig:Om_h_margw0},
\ref{fig:Om_h_margw0w1} and \ref{fig:Om_h_margw0w1zast}.  However,
when all the data are combined, there is little
difference between the constraints for the cases with a non-flat
universe being allowed and a cosmological constant with a flat
universe. This is because the combination of all three cosmological data sets 
work well to remove the degeneracy even if we consider the possibilities of a
non-flat universe. 
On the other hand, when one focuses on the combination
of two data sets such as CMB+BAO and CMB+SN, we can see some
differences between a flat case and a non-flat case.  Recall that when
one assumes a flat universe, the combination of CMB and SN already
gives a relatively severe constraint even for the case with
time-varying equations of state but that of CMB+BAO is not so severe.
For the case with a non-flat universe, in contrast to the flat universe case,
the combination of CMB+SN does not give a severe constraint.
When one allows the nonzero curvature of the
universe, the degeneracy in CMB (here this means $\theta_A$) becomes
worse: it involves $\Omega_m, \Omega_k$ and dark energy
parameters. Thus just adding SN data is not enough to break this
degeneracy. BAO is necessary to remove such degeneracy.  In Appendix
\ref{sec:OmOk}, we discuss this point in a quantitative way.

We have seen that the distance measures suffer from the degeneracies
among parameters describing dark energy equation of state and the
curvature of the universe.  Due to those degeneracies, if we use CMB
alone, CMB+BAO and CMB+SN, we get different constraints as clearly
seen in Figs.~\ref{fig:Om_h_margcurv_lambda} and
\ref{fig:Om_h_margcurv_w0}.  It seems to be necessary to combine all
three observations to remove the degeneracies.  Once that is done, the
constraints are almost the same as the case with the flat $\Lambda$CDM
model.  In Table~\ref{tab:margDE_Ok}, the central values and 1$\sigma$
errors of $h$ and $\Omega_m$ are summarized for the cases with a
cosmological constant, a constant equation of state, dark energy
parametrizations given in Eqs.~\eqref{eq:eos} and \eqref{eq:eos2} when
a non-flat universe is allowed in the analysis.  The table shows that
the constraint on $h$ is almost unchanged even if we allow a non-flat
universe under some types of dark energy parametrization.

\subsubsection{Summary of constraints on the Hubble constant} \label{subsubsec:summary_h}
\begin{figure}
	\vspace{-2cm}
    \centerline{\scalebox{1}{\includegraphics{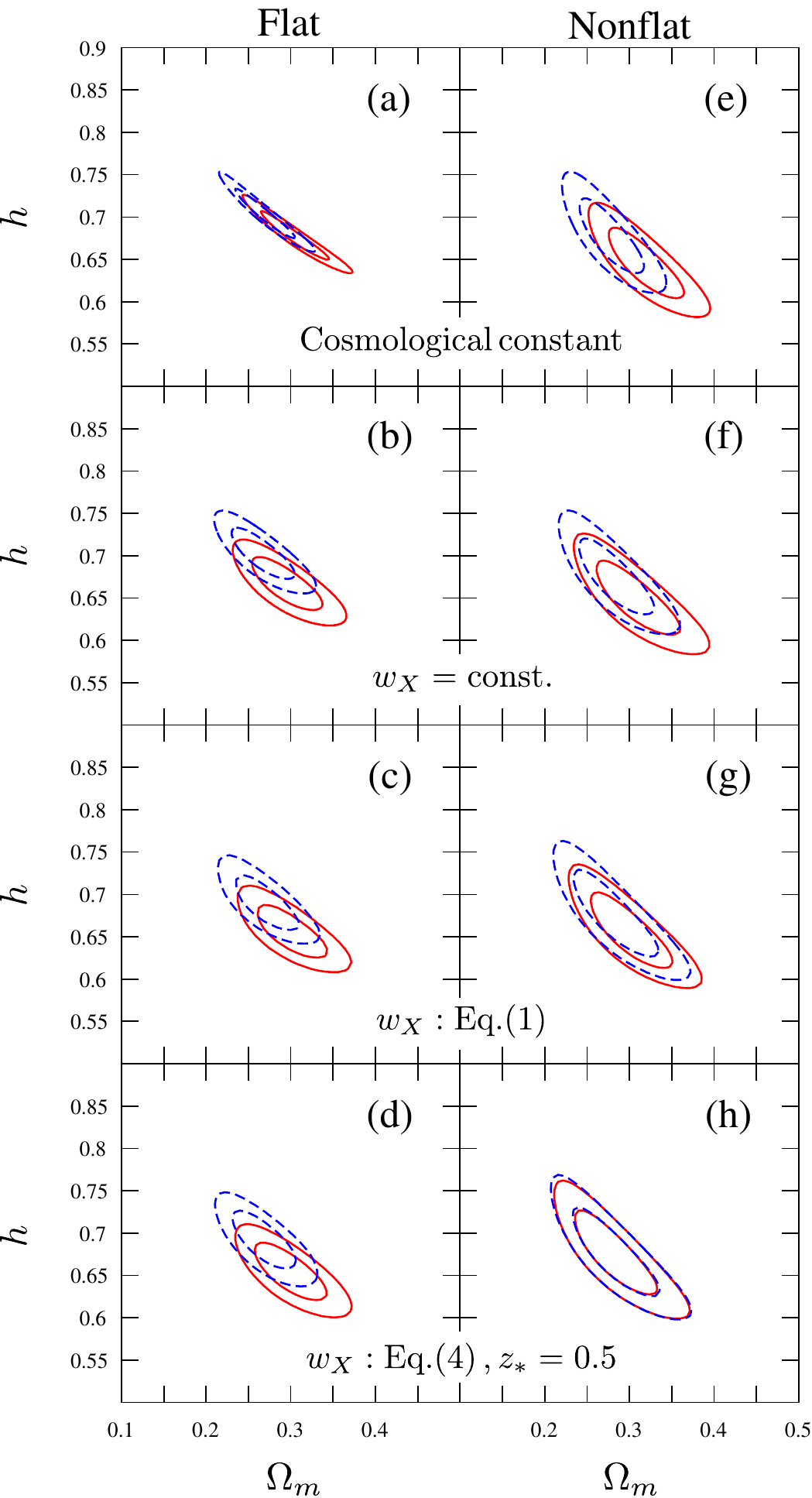}}}
    \caption{Contours 1$\sigma$ and 2$\sigma$ allowed regions in
      the $\Omega_m$--$h$ plane for the cases with: (a) a cosmological constant (same as
      Fig.~\ref{fig:Om_h_cc} (d)), (b) a constant equation of state $w_X$,
      marginalized over $w_X$, (c) the parametrization defined by Eq.~(\ref{eq:eos}),
      marginalized over $w_0$ and $w_1$, and (d) the parametrization defined by
      Eq.~(\ref{eq:eos2}) with $z_\ast = 0.5$, marginalized over
      $\tilde{w}_0$ and $\tilde{w}_1$. Constraints using the SN data
      from Gold06 (red solid line) and Davis07 (blue dashed line) are
      shown separately. In the panels (e)--(h), we marginalize over $\Omega_k$ in
      addition to the panels (a)--(d) respectively.}
	\label{fig:Om_h_margDE}
\end{figure}

In Fig.~\ref{fig:Om_h_margDE}, we summarize some of the results for
the constraints on $h$ and $\Omega_m$ presented in this section. See
also Tables~\ref{tab:margDE} and \ref{tab:margDE_Ok} for more detailed
constraints on $h$ and $\Omega_m$ and the best fit values of dark
energy parameters and/or the curvature.  As discussed above, even if
we assume different types of dark energy parametrization, there is no
considerable change in the constraints on $h$ and $\Omega_m$, although
the allowed region for the case with a cosmological constant and a
flat universe is slightly smaller compared to the other cases. 
It is interesting to compare this result with the distance ladder measurements of $h$
(Sec.~\ref{subsec:method_H}).
In particular, when all the data are combined, the central values of $h$
cannot be as low as the Sandage's central value 0.62 and cannot be as
high as the Macri's central value 0.74 even if we relax the
assumptions of a cosmological constant as dark energy and/or the
flatness of the universe. Of course, since the current measurements of $h$ by
the distance ladder have somewhat large systematic errors (see Table
\ref{tab:summary_h}), this must be taken only as a quick
comparison. 
The most seemingly problematic case arises when we compare CMB+BAO+Davis07 and Sandage's $h$,
but their 1\,$\sigma$ errors overlap.
Our analysis shows that the allowed region of the Hubble
constant from CMB, BAO and SN can be different from the WMAP flat
$\Lambda$CDM value of $h = 0.73 \pm 0.03$ by more than 1$\sigma$ 
depending on the assumption for dark energy and the cosmic curvature, 
but $h < 0.59$ or $h > 0.76$ are not allowed at 2$\sigma$ level which can
be read off from Tables~\ref{tab:margDE} and \ref{tab:margDE_Ok}. This
conclusion is obtained for rather limited types of dark energy
parametrization but the interpretation presented in the appendix leads us
to speculate that this holds true for any dark energy parametrization.
  
It should also be noticed that 
the choice of the SN data set makes larger difference than that of 
the assumptions on dark energy
and the curvature of the universe. 
The Gold06 data tends to give lower
$h$ and higher $\Omega_m$ than the Davis07 data.
Thus, there is much room for improvement in SN data as well as the distance ladder measurement of $h$.
We need more precise measurement from both fields of observation to tell whether there is a discrepancy or not.

The errors on the Hubble constant (and on $\Omega_m$) are slightly
larger for the cases with a non-flat universe than those with a flat
universe which can be seen from Tables~\ref{tab:margDE} and
\ref{tab:margDE_Ok}.  For a flat case, 1$\sigma$ errors on $h$ are about
$0.02$ and, for a non-flat case, it is around $0.03$. 
Therefore, when we use the current observations of CMB, BAO and SN,
we can expect that this level of accuracy is needed for the distance
ladder measurement of the Hubble constant to give a meaningful
external prior on $h$ in constraining the dark energy parameters and
the curvature of the universe.  Assuming that such sensitivity on $h$
is achieved in the future, we investigate its effects on the determination
of dark energy equation of state and the curvature of the universe in
the following sections \ref{subsec:constraint_DE} and
\ref{subsec:constraint_Ok}.

\subsection{Constraint on dark energy equations of state} \label{subsec:constraint_DE}

Here we discuss the constraints on parameters of dark energy which
describe the time dependence of its equation of state.  Our special
emphasis is on investigating how they are affected by several external
priors on the Hubble constant and/or by the assumption of the flat
universe. 
In most analyses on this issue so far,  a flat universe is usually assumed.
Once one invokes the inflationary paradigm, the flatness assumption 
seems to be natural. However, the flatness itself should be tested 
including the uncertainties of dark energy. 
In this respect, when we investigate the equation of state
for dark energy, we also remove the assumption of a flat universe to 
obtain a more conservative constraint. 
In particular, in the following, we discuss
implications of the results taking a cosmological constant
as a reference point. 
Since a cosmological constant is the simplest and most conventional 
model for dark energy (and it can fit the observations satisfactorily as shown below),
 it is useful for illustrating the roles of the priors.

\begin{figure}
    \centerline{\scalebox{1}{\includegraphics{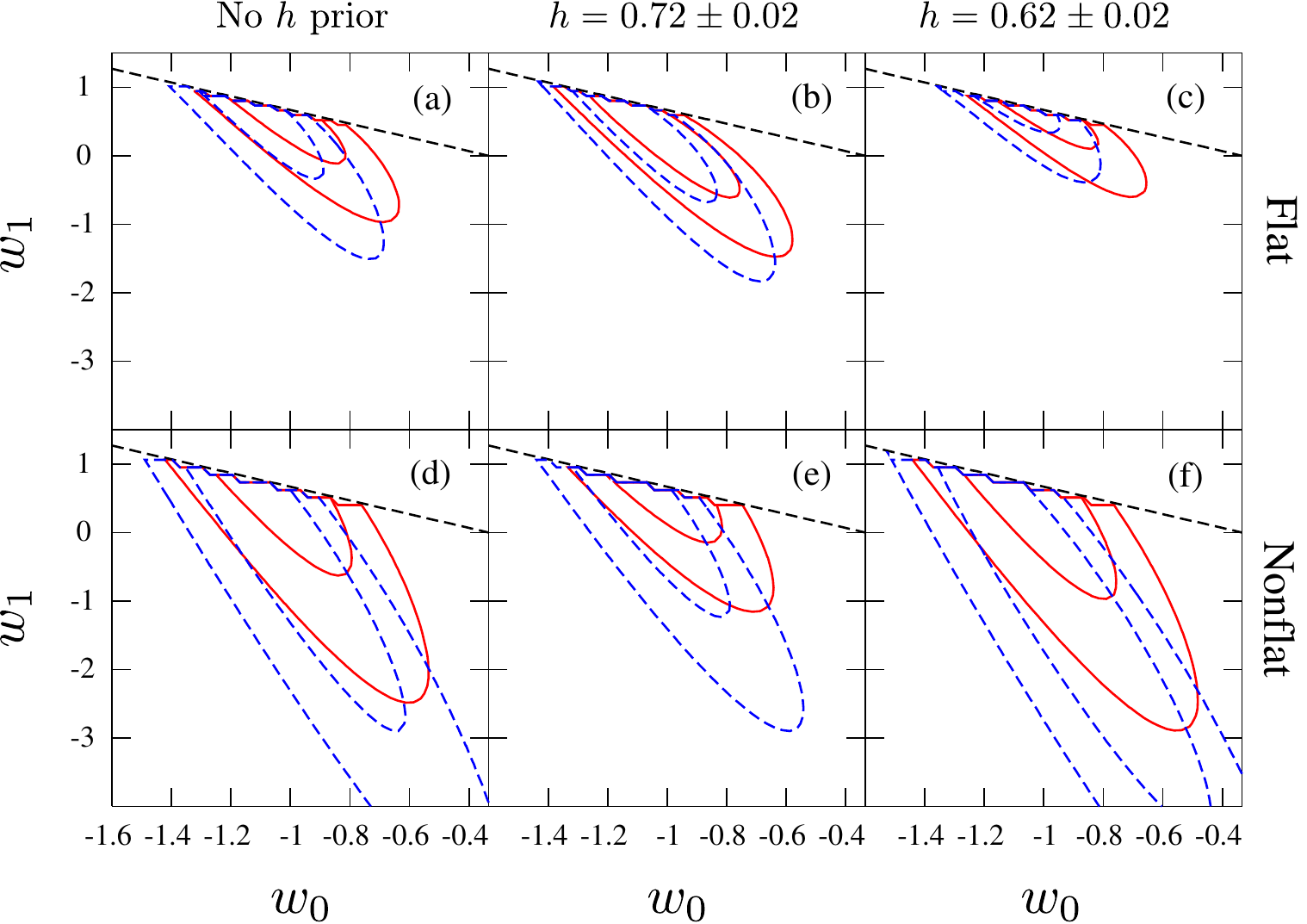}}}
    \caption{1$\sigma$ and 2$\sigma$ constraints from CMB+BAO+SN in
      the $w_0$--$w_1$ plane marginalizing over $\Omega_m$ and $h$ are
      shown for the cases with (a) no prior on the Hubble constant,
      (b) assuming a Gaussian prior on the Hubble constant $h = 0.72
      \pm 0.02$ and (c) $h = 0.62 \pm 0.02$. In the panels (d)--(f),
      we allow a non-flat universe and marginalize over $\Omega_k$ in
      addition to $\Omega_m$ and $h$.  The black dashed lines show
      the boundary of the prior Eq.~\eqref{eq:eos_prior2}.  The
      constraints using the SN data sets from Gold06 (red solid line)
      and Davis07 (blue dashed line) are shown separately.}
	\label{fig:w0_w1}
\end{figure}

First we assume the dark energy parametrization of Eq.~\eqref{eq:eos}.
In Fig.~\ref{fig:w0_w1}, the constraints in the $w_0$--$w_1$ plane are
shown for several cases.  In the panels (a)-(c), we marginalized over the
values of $\Omega_m$ and $h$.  In the panels (d)-(f), we repeat the same
analysis except that we allow a non-flat universe and marginalize over
$\Omega_k$.  The best fit values of $w_0$ and $w_1$ and the minimum
$\chi^2$ are summarized in Tables~\ref{tab:w0w1_flat} and
\ref{tab:w0w1_nonflat}. The best fit values of $h$ and $\Omega_m$ (and
$\Omega_k$ for the case of a non-flat universe in Table
\ref{tab:w0w1_nonflat}) are also shown.  In the panels (b), (c), (e)
and (f), we assume some priors on the Hubble constant to see how it can
affect the determination of dark energy parameters.  We impose
Gaussian priors as $h = 0.72 \pm 0.02$ (panels (b) and (e)) and $h =
0.62 \pm 0.02$ (panels (c) and (f)) where we use the Freedman's and
Sandage's (see Table~\ref{tab:summary_h}) as the central values and
hypothetical errors of 0.02.

The hypothetical errors adopted here are motivated from the results we
obtained in Sec.~\ref{subsec:constraint_H}. As summarized in
Sec.~\ref{subsubsec:summary_h}, the uncertainties of $h$ from the
cosmological observations are $0.02 \sim 0.03$.  Thus we can expect
that this level of accuracy on the Hubble constant prior is required
to have an influence on constraining dark energy sector.  In fact, we
have also made the analysis adopting a Gaussian prior with the current
errors as shown in Table~\ref{tab:summary_h} but we cannot see a
noticeable difference compared with the case of no prior on the Hubble constant.
It should be noted that, as mentioned in Sec.~\ref{subsec:method_H}, this
level of improvement on the Hubble constant determination is not
unimaginable in near future \cite{Macri:2006wm}.  Note that we can
expect that other observations (CMB, BAO and SN) would be more precise
at the time when this level of accuracy in the Hubble constant
determination is realized. Thus, our combining the hypothetical Hubble
priors and the current observational data is not a forecast of future
status in a strict sense but should rather be regarded as an
illustration of how the Hubble external priors affect conclusions on
properties of dark energy and the curvature of the universe.
 
 Let us start with looking
at the constraints from the present cosmological observations assuming
no Hubble prior.  When we assume a flat universe, although the best
fit value is slightly away from a cosmological constant (especially in
the positive direction of $w_1$. See Table~\ref{tab:w0w1_flat}), a
cosmological constant is within the 2$\sigma$ allowed regions for both
SN data sets as shown in Fig.~\ref{fig:w0_w1} (a). This situation
holds when we allow a non-flat universe and marginalize over
$\Omega_k$ as shown in Fig.~\ref{fig:w0_w1} (d). We can understand
this by noticing that the region closely around a flat universe is
favored in this parametrization with no Hubble prior as seen in Table
\ref{tab:w0w1_nonflat} or in the analysis presented in the following
Sec.~\ref{subsec:constraint_Ok}.

\begin{table}
  \centering 
  \begin{tabular}{|l||c|c|c|c|c|}
  \hline
  CMB+BAO+Gold06 & $\chi^2_{\rm min}$ & $\Omega_m$ & $h$ & $w_0$ & $w_1$ \\
  \hline
  No prior  & 158.3 & 0.301 & 0.654 & -1.06 & 0.72 \\
  Prior $h=0.72 \pm 0.02$ & 163.1 & 0.267 & 0.692 & -1.03 & 0.43 \\
  Prior $h=0.62 \pm 0.02$ & 159.6 & 0.317 & 0.640 & -1.02 & 0.68 \\
   \hline
  \hline
 CMB+BAO+Davis07 & $\chi^2_{\rm min}$ & $\Omega_m$ & $h$ & $w_0$ & $w_1$ \\
  \hline
 No prior     &  195.5 & 0.272 & 0.686 & -1.16 & 0.82 \\
  Prior $h=0.72 \pm 0.02$ & 196.8  & 0.256  & 0.707 & -1.10 & 0.52 \\
   Prior $h=0.62 \pm 0.02$ & 200.9  & 0.301 & 0.658 & -1.10 & 0.77 \\
   \hline
   \end{tabular}
   \caption{The best fit values for $\Omega_m$, $h$, $w_0$ and $w_1$ for 
     the analysis presented in the panel (a)-(c) of Fig.~\ref{fig:w0_w1}.
     The minimum values of $\chi^2$ are also shown.}
   \label{tab:w0w1_flat}
  \end{table}
\begin{table}
  \centering 
  \begin{tabular}{|l||c|c|c|c|c|c|}
  \hline
  CMB+BAO+Gold06 & $\chi^2_{\rm min}$ & $\Omega_m$ & $h$ & $\Omega_k$ & $w_0$ & $w_1$ \\
  \hline
   No prior & 158.3 & 0.301 & 0.655 & 0.000 & -1.06 & 0.72 \\
   prior $h=0.72 \pm 0.02$ & 161.3 & 0.263 & 0.702 & 0.014 & -1.02 & 0.69 \\
  Prior $h=0.62 \pm 0.02$ & 159.2 & 0.320 & 0.634 & -0.007 & -1.06 & 0.73 \\
   \hline
  \hline
 CMB+BAO+Davis07 & $\chi^2_{\rm min}$ & $\Omega_m$ & $h$ & $\Omega_k$ & $w_0$ & $w_1$ \\
  \hline
   No prior &  195.4 & 0.279 & 0.676 & -0.009 & -1.10 & 0.39 \\
  Prior $h=0.72 \pm 0.02$ & 196.6  & 0.254  & 0.709 & 0.006 & -1.14 & 0.80 \\
  Prior $h=0.62 \pm 0.02$ & 197.3  & 0.315 & 0.638 & -0.026 & -0.93 & -0.92 \\
   \hline
   \end{tabular}
   \caption{The best fit values for $\Omega_m$, $h$, 
     $\Omega_k$, $w_0$ and $w_1$ for the analysis presented 
     in the panel (d)-(f) of Fig.~\ref{fig:w0_w1}.
     The minimum values of $\chi^2$ are also shown.}
   \label{tab:w0w1_nonflat}
\end{table}
  
We now turn to the constraints when we assume the external priors on the Hubble
constant. The constraints in the
$w_0$--$w_1$ plane for this case are shown in Fig.~\ref{fig:w0_w1} (b), (c), (e) and (f).
Probably the most interesting feature is seen in the panel (c): in a
flat universe, if the Sandage's value is confirmed at this level, a
cosmological constant ($w_0 = -1$ and $w_1 = 0$ in the figure) would
be rejected at nearly 2$\sigma$ level in this parametrization.
However, as shown in the panel (f), if we allow a non-flat universe
and marginalize over $\Omega_k$, a cosmological constant is well
within the allowed regions.  This is an example which shows the
importance of the priors on the Hubble constant and the curvature of
the universe when we probe the nature of dark energy.

\begin{figure}
    \centerline{\scalebox{0.9}{\includegraphics{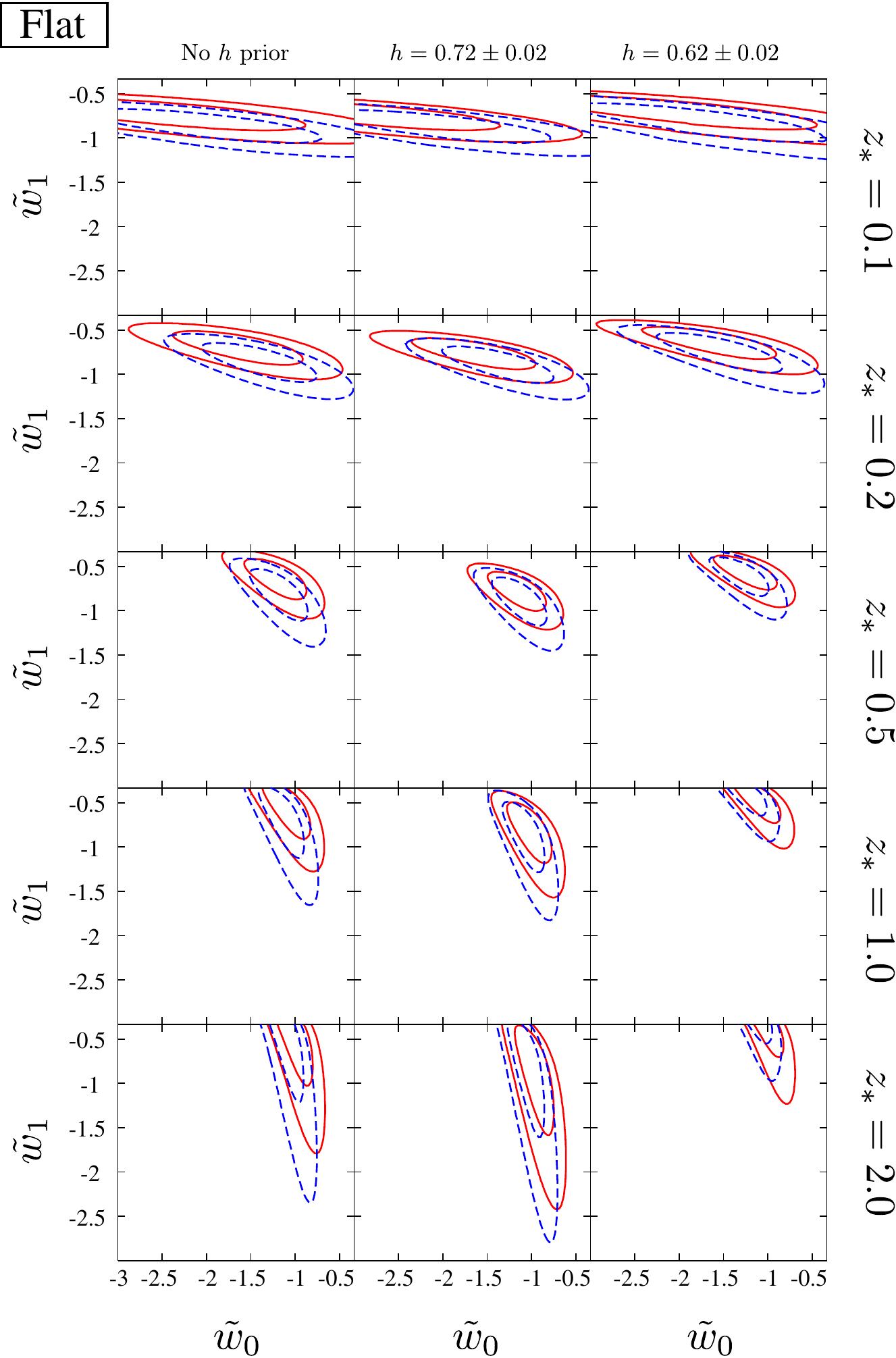}}}
    \caption{1$\sigma$ and 2$\sigma$ constraints from CMB+BAO+SN in
      the $\tilde{w}_0$--$\tilde{w}_1$ plane marginalizing over $h$
      and $\Omega_m$ are shown for several values of $z_\ast$. For
      each $z_\ast$, we show the cases with no prior on the Hubble
      constant and Gaussian priors $h = 0.72
      \pm 0.02$ and $h = 0.62 \pm 0.02$. A flat universe is assumed.
      The constraints using the SN data sets from Gold06 (red solid
      line) and Davis07 (blue dashed line) are shown separately.}
      \label{fig:w0w1_zast}
\end{figure}

\begin{figure}
    \centerline{\scalebox{0.9}{\includegraphics{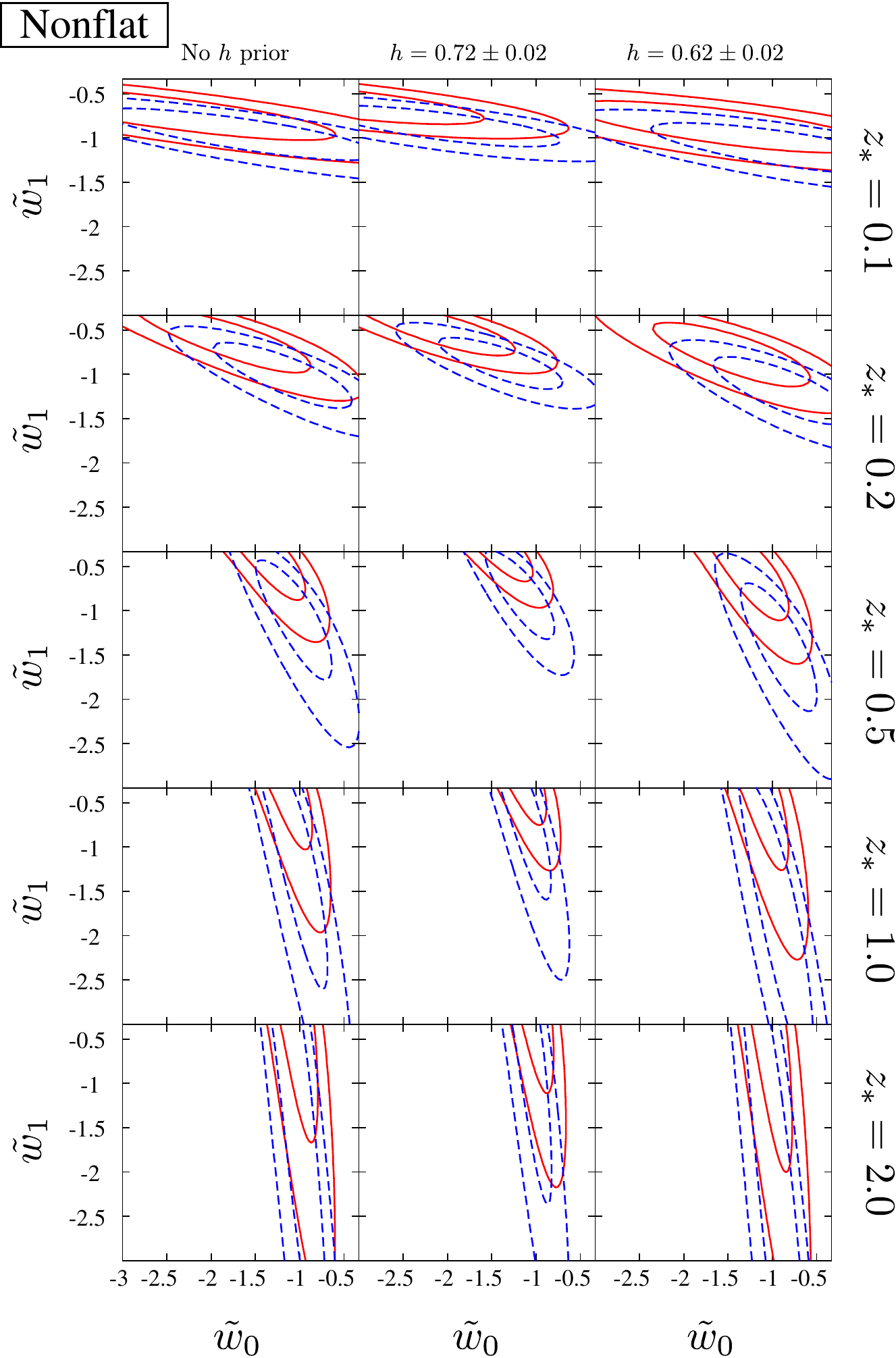}}}
    \caption{1$\sigma$ and 2$\sigma$ constraints from CMB+BAO+SN in
      the $\tilde{w}_0$--$\tilde{w}_1$ plane marginalizing over $h$,
      $\Omega_m$ and $\Omega_k$ are shown for several values of
      $z_\ast$ for the cases with no Hubble prior and assuming Gaussian
      priors $h = 0.72 \pm 0.02$ and $h = 0.62
      \pm 0.02$.}
          \label{fig:w0w1_zast_Okmarg}
\end{figure}

\begin{table}
  \centering 
  \begin{tabular}{|l||c|c|c|c|c|}
  \hline
  CMB+BAO+Gold06 & $\chi^2_{\rm min}$ & $\Omega_m$ & $h$ & $\tilde{w}_0$ & $\tilde{w}_1$ \\
  \hline
 $z_\ast = 0.1$ (No prior)  & 157.9 & 0.279 & 0.678 & -2.52 & -0.71 \\
  Prior $h=0.72 \pm 0.02$ & 159.8 & 0.257 & 0.702 & -2.83 & -0.73 \\
  Prior $h=0.62 \pm 0.02$ & 161.2 & 0.307 & 0.648 & -2.11 & -0.68 \\
\hline
$z_\ast = 0.2$ (No prior)  & 157.5 & 0.289 & 0.665 & -1.63 & -0.68 \\
  Prior $h=0.72 \pm 0.02$ & 161.0 & 0.262 & 0.695 & -1.63 & -0.75 \\
  Prior $h=0.62 \pm 0.02$ & 159.6 & 0.310 & 0.643 & -1.65 & -0.62 \\
\hline
$z_\ast = 0.5$ (No prior)  & 157.5 & 0.298 & 0.654 & -1.21 & -0.61 \\
  Prior $h=0.72 \pm 0.02$ & 162.2 & 0.265 & 0.692 & -1.15 & -0.75 \\
  Prior $h=0.62 \pm 0.02$ & 158.6 & 0.315 & 0.637 & -1.24 & -0.54 \\
\hline
$z_\ast = 1.0$ (No prior)  & 158.1 & 0.304 & 0.649 & -1.08 & -0.52 \\
  Prior $h=0.72 \pm 0.02$ & 163.1 & 0.266 & 0.692 & -1.01 & -0.77 \\
  Prior $h=0.62 \pm 0.02$ & 158.8 & 0.319 & 0.634 & -1.11 & -0.44 \\
\hline
$z_\ast = 2.0$ (No prior)  & 158.6 & 0.306 & 0.649 & -1.02 & -0.38 \\
  Prior $h=0.72 \pm 0.02$ & 163.4 & 0.266 & 0.693 & -0.96 & -0.77 \\
  Prior $h=0.62 \pm 0.02$ & 159.3 & 0.319 & 0.635 & -1.01 & -0.33 \\
   \hline
  \hline
 CMB+BAO+Davis07 & $\chi^2_{\rm min}$ & $\Omega_m$ & $h$ 
 & $\tilde{w}_0$ & $\tilde{w}_1$ \\
  \hline
$z_\ast = 0.1$ (No prior)     &  195.5 & 0.258 & 0.707 & -2.00 & -0.85 \\
  Prior $h=0.72 \pm 0.02$ & 195.7  & 0.251  & 0.714 & -2.07 & -0.85 \\
   Prior $h=0.62 \pm 0.02$ & 204.1  & 0.297 & 0.665 & -1.76 & -0.80 \\
   \hline
 $z_\ast = 0.2$ (No prior)     &  195.9 & 0.263 & 0.700 & -1.40 & -0.84 \\
  Prior $h=0.72 \pm 0.02$ & 196.4  & 0.253  & 0.710 & -1.39 & -0.86 \\
   Prior $h=0.62 \pm 0.02$ & 203.2  & 0.297 & 0.662 & -1.54 & -0.72 \\
   \hline
$z_\ast = 0.5$ (No prior)     &  195.9 & 0.267 & 0.692 & -1.18 & -0.77 \\
  Prior $h=0.72 \pm 0.02$ & 196.7  & 0.255  & 0.708 & -1.14 & -0.84 \\
   Prior $h=0.62 \pm 0.02$ & 201.2  & 0.298 & 0.653 & -1.31 & -0.57 \\
   \hline
$z_\ast = 1.0$ (No prior)     &  195.5 & 0.273 & 0.684 & -1.15 & -0.61 \\
  Prior $h=0.72 \pm 0.02$ & 196.8  & 0.256  & 0.706 & -1.09 & -0.77 \\
   Prior $h=0.62 \pm 0.02$ & 199.0  & 0.303 & 0.646 & -1.24 & -0.38 \\
   \hline
$z_\ast = 2.0$ (No prior)     &  195.4 & 0.277 & 0.678 & -1.12 & -0.38 \\
  Prior $h=0.72 \pm 0.02$ & 196.9  & 0.257  & 0.706 & -1.06 & -0.67 \\
   Prior $h=0.62 \pm 0.02$ & 199.3  & 0.301 & 0.653 & -1.08 & -0.33 \\
   \hline
\end{tabular}
\caption{The best fit values for $\Omega_m$, $h$, $\tilde{w}_0$ and 
  $\tilde{w}_1$ for the analysis presented in  Fig.~\ref{fig:w0w1_zast} 
  in which a flat universe is assumed.
  The minimum values of $\chi^2$ are also shown.}
    \label{tab:w0w1_zast}
\end{table}

\begin{table}
  \centering 
  \begin{tabular}{|l||c|c|c|c|c|c|}
  \hline
  CMB+BAO+Gold06 & $\chi^2_{\rm min}$ & $\Omega_m$ & $h$ & $\Omega_k$ & $\tilde{w}_0$ & $\tilde{w}_1$ \\
  \hline
 $z_\ast = 0.1$ (No prior)  & 157.8 & 0.273 & 0.684 & 0.003 & -2.66 & -0.68 \\
  Prior $h=0.72 \pm 0.02$ & 158.4 & 0.251 & 0.714 & 0.013 & -3.20 & -0.60 \\
  Prior $h=0.62 \pm 0.02$ & 159.7 & 0.318 & 0.636 & -0.013 & -1.58 & -0.85 \\
\hline
$z_\ast = 0.2$ (No prior)  & 157.3 & 0.278 & 0.678 & 0.009 & -1.80 & -0.59 \\
  Prior $h=0.72 \pm 0.02$ & 158.3 & 0.253 & 0.711 & 0.025 & -2.06 & -0.49 \\
  Prior $h=0.62 \pm 0.02$ & 159.2 & 0.316 & 0.637 & -0.010 & -1.44 & -0.75 \\
\hline
$z_\ast = 0.5$ (No prior)  & 156.6 & 0.279 & 0.676 & 0.032 & -1.39 & -0.33 \\
  Prior $h=0.72 \pm 0.02$ & 157.9 & 0.255 & 0.709 & 0.042 & -1.38 & -0.33 \\
  Prior $h=0.62 \pm 0.02$ & 158.6 & 0.314 & 0.638 & 0.002 & -1.25 & -0.51 \\
\hline
$z_\ast = 1.0$ (No prior)  & 157.5 & 0.294 & 0.661 & 0.018 & -1.12 & -0.33 \\
  Prior $h=0.72 \pm 0.02$ & 160.0 & 0.260 & 0.704 & 0.032 & -1.09 & -0.33 \\
  Prior $h=0.62 \pm 0.02$ & 158.7 & 0.318 & 0.635 & 0.009 & -1.14 & -0.33 \\
\hline
$z_\ast = 2.0$ (No prior)  & 158.5 & 0.302 & 0.653 & 0.005 & -1.01 & -0.33 \\
  Prior $h=0.72 \pm 0.02$ & 161.7 & 0.263 & 0.702 & 0.020 & -0.98 & -0.33 \\
  Prior $h=0.62 \pm 0.02$ & 159.3 & 0.321 & 0.633 & -0.003 & -1.03 & -0.33 \\
   \hline
  \hline
 CMB+BAO+Davis07 & $\chi^2_{\rm min}$ & $\Omega_m$ & $h$ & $\Omega_k$ & $\tilde{w}_0$ & $\tilde{w}_1$ \\
  \hline
$z_\ast = 0.1$ (No prior)     &  195.1 & 0.271 & 0.686 & -0.008 & -1.70 & -0.93 \\
  Prior $h=0.72 \pm 0.02$ & 195.7  & 0.251  & 0.713 & -0.001 & -2.04 & -0.86 \\
   Prior $h=0.62 \pm 0.02$ & 197.5  & 0.312 & 0.639 & -0.023 & -0.94 & -1.09 \\
   \hline
 $z_\ast = 0.2$ (No prior)     &  195.4 & 0.278 & 0.678 & -0.010 & -1.20 & -0.97 \\
  Prior $h=0.72 \pm 0.02$ & 196.4  & 0.253  & 0.711 & 0.000 & -1.38 & -0.86 \\
   Prior $h=0.62 \pm 0.02$ & 197.4  & 0.313 & 0.638 & -0.023 & -0.94 & -1.14 \\
   \hline
$z_\ast = 0.5$ (No prior)     &  195.5 & 0.281 & 0.675 & -0.011 & -1.07 & -0.98 \\
  Prior $h=0.72 \pm 0.02$ & 196.7  & 0.254  & 0.710 & 0.002 & -1.16 & -0.79 \\
   Prior $h=0.62 \pm 0.02$ & 197.3  & 0.315 & 0.637 & -0.026 & -0.93 & -1.29 \\
   \hline
$z_\ast = 1.0$ (No prior)     &  195.4 & 0.279 & 0.677 & -0.008 & -1.09 & -0.84 \\
  Prior $h=0.72 \pm 0.02$ & 196.5  & 0.253  & 0.710 & 0.013 & -1.17 & -0.47 \\
   Prior $h=0.62 \pm 0.02$ & 197.3  & 0.315 & 0.637 & -0.026 & -0.96 & -1.52 \\
   \hline
$z_\ast = 2.0$ (No prior)     &  195.4 & 0.278 & 0.677 & -0.005 & -1.10 & -0.53 \\
  Prior $h=0.72 \pm 0.02$ & 196.6  & 0.255  & 0.709 & 0.011 & -1.10 & -0.33 \\
   Prior $h=0.62 \pm 0.02$ & 197.3  & 0.315 & 0.637 & -0.027 & -0.96  & -2.08 \\
   \hline
\end{tabular}
\caption{The best fit values for $\Omega_m$, $h$, $\Omega_k$, $\tilde{w}_0$ and 
  $\tilde{w}_1$ for 
  the analysis presented in  Fig.~\ref{fig:w0w1_zast_Okmarg} in which 
  a non-flat  universe is allowed.
  The minimum values of $\chi^2$ are also shown.}
    \label{tab:w0w1_zast_Okmarg}
\end{table}

Next we show the constraints on the dark energy
parameters for the parametrization Eq.~\eqref{eq:eos2} in Figs.~\ref{fig:w0w1_zast} and
\ref{fig:w0w1_zast_Okmarg}.  Although this
parametrization includes three parameters, we present our result in
the $\tilde{w}_0$--$\tilde{w}_1$ plane fixing $z_\ast$ to several
values: 0.1, 0.2, 0.5, 1.0 and 2.0.  Namely we regard $z_\ast$ as
labeling the model expressed as Eq.~\eqref{eq:eos2} which has two
parameters $\tilde{w}_0$ and $\tilde{w}_1$, and do not marginalize
over $z_\ast$. In Fig.~\ref{fig:w0w1_zast}, a flat universe is assumed
and in Fig.~\ref{fig:w0w1_zast_Okmarg}, we allow a non-flat universe
and marginalize over the curvature. The best fit parameter values
are shown respectively in Tables~\ref{tab:w0w1_zast} and
\ref{tab:w0w1_zast_Okmarg}.  We also impose some priors on the Hubble
constant as is done for the analysis of the dark energy
parametrization of Eq.~\eqref {eq:eos}.
  
For the case with no Hubble prior,
as the panels in the left column in Figs.~\ref{fig:w0w1_zast} and \ref{fig:w0w1_zast_Okmarg} indicate,
a cosmological constant ($\tilde{w}_0 = -1$ and $\tilde{w}_1=-1$) is
in the allowed regions with or without the assumption of a flat
universe. This is because, as is the case with the parametrization
Eq.~\eqref{eq:eos} discussed above, the region around the flat
universe is more or less favored in this parametrization with no
Hubble prior as seen in Tables~\ref{tab:w0w1_zast} and
\ref{tab:w0w1_zast_Okmarg}.
  
However, if we impose some external priors on the Hubble constant, 
the constraints in the $ \tilde{w}_0$--$\tilde{w}_1$ plane receive some interesting effects
 of the external priors on the Hubble
constant and the assumption for the curvature of the universe.
We can see them from the panels in the middle and right columns in
Figs.~\ref{fig:w0w1_zast} and \ref{fig:w0w1_zast_Okmarg}. 

For example, when we impose $h=0.62 \pm 0.02$ for the $z_\ast = 1.0$ model assuming
a flat universe, a cosmological constant is not within the 2$\sigma$
allowed region for both SN data sets (see Fig.~\ref
{fig:w0w1_zast}). However, if we allow a non-flat universe and
marginalize over $\Omega_k$, a cosmological constant is allowed with
2$\sigma$ level  (see Fig.~\ref{fig:w0w1_zast_Okmarg}).
This is similar to the tendency which is seen in the
case of the parametrization Eq.~\eqref{eq:eos} discussed
above. Another interesting point can be observed in this
parametrization, which is actually an opposite tendency to the one
just mentioned. When we impose $h=0.72 \pm 0.02$ for the $z_\ast
= 0.5$ model assuming a flat universe, a cosmological constant is well
within the allowed regions for both SN data sets (see
Fig.~\ref{fig:w0w1_zast}). However, when we use the Gold06 data set,
if we allow a non-flat universe and marginalize over $\Omega_k$, a
cosmological constant is out of the 2$\sigma$ allowed region (see
Fig.~\ref{fig:w0w1_zast_Okmarg}). This is considered to be an
infrequent case in which relatively large $\Omega_k$ gives a
significantly better fit than a flat universe ($\chi^2_{\rm min}$
decreases by 4.3 in this case) with the equation of state well away
from a cosmological constant, but we should bear in mind that such a
case could happen.
  
In summary, we have seen that the prior on the Hubble constant, the
way to parametrize the equation of state and the assumption for the
curvature of the universe are all very important to probe the nature
of dark energy. 
These would be true for data from future experiments of CMB, BAO and SN.
In particular, since there is a well-known severe degeneracy among $w_X$ and $\Omega_X$,
the independent determination of the Hubble constant will still affect 
the constraint on dark energy parameters by pinning down the value of 
$\Omega_m$ more precisely. 
The answer to the simplest question of ``a
cosmological constant or not" can be altered by changing some of these
assumptions.

\subsection{Constraint on the curvature of the universe} \label{subsec:constraint_Ok}

Here we discuss the constraints on the curvature of the universe
assuming some priors on the Hubble constant including some types of
dark energy model in addition to a cosmological constant and a
constant equation of state using the parametrizations
Eqs.~\eqref{eq:eos} and \eqref{eq:eos2}.  As already mentioned, a flat
universe is assumed in most cosmological parameter estimation in the literature
because the inflationary paradigm strongly suggests it. 
However, the paradigm should be tested through the test of the flatness and it should be
done including the uncertainty in dark energy sector since its nature is not understood yet. 
In this respect, the
investigation of the curvature of the universe in connection with dark
energy models have been done in
Refs.~\cite{Crooks:2003pa,Balbi:2003en,Linder:2005nh,Dick:2006ev,Huang:2006er,Knox:2006ux,Smith:2006nk,Hu:2006nm,Zhao:2006qg,Ichikawa:
  2005nb,Ichikawa:2006qb,Ichikawa:2006qt,Polarski:2005jr,Knox:2005hx,Gong:2007wx,Gong:2005de,Clarkson:2007bc,Rapetti:2007mw}.
In Refs.~\cite{Ichikawa:2005nb,Ichikawa:2006qb,Ichikawa:2006qt}, it
was shown that an open universe can be largely allowed for some
particular dark energy parametrizations. However, in the previous
works, the Hubble constant was treated as a nuisance parameter to be
marginalized over.  Here we analyze how the constraint on the
curvature is correlated with the Hubble constant and how the priors on
it affect the constraint.  We also investigate whether the constraint
varies according to the choice of dark energy parametrization.
 
  \begin{figure}
 \vspace{-3cm}
    \centerline{\scalebox{0.9}{\includegraphics{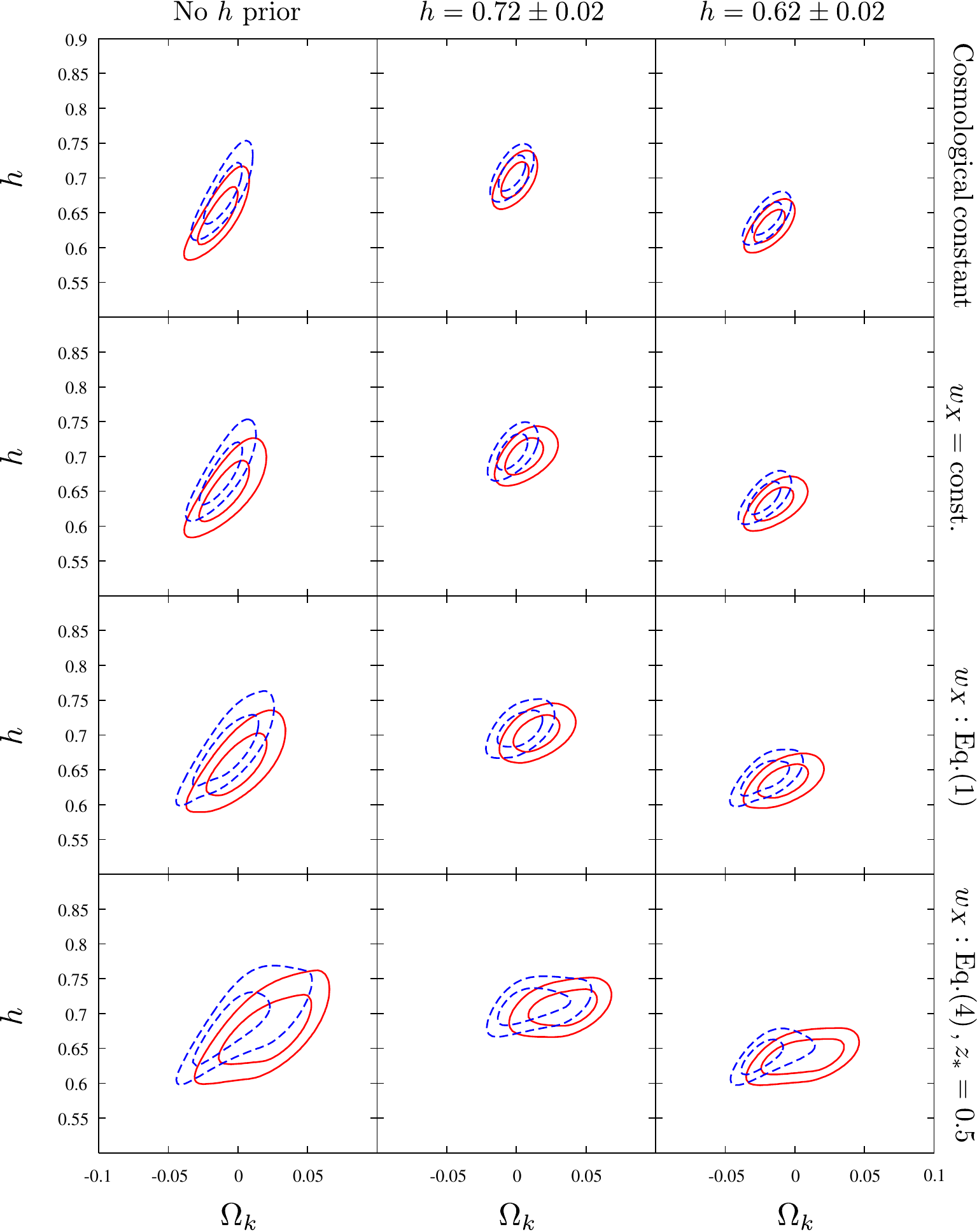}}}
    \caption{1$\sigma$ and 2$\sigma$ constraints from CMB+BAO+SN in
      the $\Omega_k$--$h$ plane marginalizing over $\Omega_m$ and dark
      energy parameters. The dark energy parametrization adopted in
      the analysis are shown on the right side of the panels. We show
      the cases with no prior on the Hubble constant and Gaussian
      priors $h = 0.72 \pm 0.02$ and $h = 0.62
      \pm 0.02$. The constraints using the SN data sets from Gold06
      (red solid line) and Davis07 (blue dashed line) are shown
      separately. }
    \label{fig:Ok_h_margOm_w0}
\end{figure}

In Fig.~\ref{fig:Ok_h_margOm_w0}, the constraints in the
$\Omega_k$--$h$ plane are shown for the cases with a cosmological
constant, a constant equation of state, time-varying equations of
state paramtrized as Eq.~\eqref{eq:eos} and Eq.~\eqref{eq:eos2} with
$z_\ast = 0.5$.  For each model, we impose no prior and two 
types of prior on the Hubble constant as in Sec.~\ref{subsec:constraint_DE}.

From the current data of CMB, BAO and SN without an external Hubble
prior, we may conclude that the universe is constrained to be around
flat. In almost all of the cases,  the curvature is limited as $|\Omega_k| <
0.05$ at 2$\sigma$ level (only exception is that the case where the
Gold06 set is used for the parametrization Eq.~\eqref{eq:eos2} with
$z_\ast = 0.5$.  The 2$\sigma$ boundary extends to $\Omega_k \sim
0.07$).  However, when we look at them more closely, we notice some
dependence on dark energy parametrization.  For the cases with a
cosmological constant and a constant equation of state, the allowed
region extends into a closed universe, which is a rather widely known
result (e.g. Fig.~17 of Ref.~\cite{Spergel:2006hy}).  In contrast, for
the case with a time-varying equation of state Eq.~\eqref{eq:eos2}
with $z_\ast = 0.5$, the allowed region has a much wider area in an
open universe as has been found in Ref.~\cite{Ichikawa:2006qb}.

Fig.~\ref{fig:Ok_h_margOm_w0} also shows the known tendency that the
lower Hubble constant is favored in a closed universe (a positive
correlation between $h$ and $\Omega_k$).  Due to this correlation,
when we assume the prior of $h= 0.62 \pm 0.02$ for the Hubble
constant, the region with a closed universe occupies larger space in
the allowed region. On the other hand, when the prior $h=0.72 \pm
0.02$ is assumed, wider region of an open universe is allowed compared
to a closed universe.  For example, a flat universe is rejected at
2$\sigma$ level and a closed universe is favored for a cosmological
constant with the prior $h= 0.62 \pm 0.02$.  Meanwhile, the prior $h=
0.72 \pm 0.02$ enhances the preference of an open universe to
2$\sigma$ level for the parametrization Eq.~\eqref{eq:eos2} with
$z_\ast = 0.5$.

\begin{table}[t]
  \centering
  \begin{tabular}{|l||c|c|c|c|c|c|}
  \hline
CMB+BAO+Gold06 & $\chi^2_{\rm min}$ & $\Omega_k$  & $h$ & $\Omega_m$ & $w_0$ ($\tilde{w}_0$) & $w_1$ ($\tilde{w}_1$) \\
  \hline
  Cosmological constant  & 159.8 & -0.014 & 0.644  & 0.318 & $-$ & $-$ \\
  Prior $h=0.72 \pm 0.02$  & 164.5 & -0.001 & 0.696  & 0.277 & $-$ & $-$ \\
  Prior $h=0.62 \pm 0.02$  & 160.2 & -0.018 & 0.630  & 0.331 & $-$ & $-$ \\
\hline
  $w_0$ ($w_1=0$) & 159.5 & -0.010 & 0.648 & 0.308 & -0.92 & $-$ \\
  Prior $h=0.72 \pm 0.02$ & 163.3 & 0.005 & 0.700 & 0.265 & -0.88 & $-$ \\
  Prior $h=0.62 \pm 0.02$ & 160.0 & -0.015 & 0.632 & 0.323 & -0.94 & $-$ \\
 \hline
  $w_0$ and $w_1$  & 158.3 & 0.000 & 0.654 & 0.301 & -1.06 & 0.72 \\
  Prior $h=0.72 \pm 0.02$  & 161.3 & 0.015 & 0.702 & 0.262 & -1.01 & 0.68 \\
 Prior $h=0.62 \pm 0.02$  & 159.2 & -0.007 & 0.634 & 0.320 & -1.07 & 0.74 \\
 \hline
  $\tilde{w}_0$ and $\tilde{w}_1$ ($z_\ast = 0.5$) & 156.6 & 0.032 & 0.676 & 0.280 & -1.38 & -0.33 \\
  Prior $h=0.72 \pm 0.02$ & 157.9 & 0.042 & 0.708 & 0.256 & -1.38 & -0.33 \\
  Prior $h=0.62 \pm 0.02$  & 158.6 & 0.002 & 0.638 & 0.313 & -1.25 & -0.52 \\
   \hline
   \hline
CMB+BAO+Davis07 & $\chi^2_{\rm min}$ & $\Omega_k$  & $h$ & $\Omega_m$ & $w_0$ ($\tilde{w}_0$) & $w_1$ ($\tilde{w}_1$) \\
  \hline
  Cosmological constant  & 195.6 & -0.011 & 0.674  & 0.278 &  $-$ & $-$ \\
  Prior $h=0.72 \pm 0.02$ & 197.1 & -0.003 & 0.706  & 0.258 &  $-$ & $-$ \\
  Prior $h=0.62 \pm 0.02$   & 198.0 & -0.020 & 0.642  & 0.302 &  $-$ & $-$ \\
\hline
 $w_0$ ($w_1=0$) & 195.5 & -0.012 & 0.674 & 0.282 & -1.04 & $-$ \\
 Prior $h=0.72 \pm 0.02$ & 197.1 & -0.004 & 0.706 & 0.258 & -1.01 & $-$ \\
 Prior $h=0.62 \pm 0.02$  & 197.6 & -0.022 & 0.640 & 0.311 & -1.07 & $-$ \\
\hline
  $w_0$ and $w_1$  &  195.4 & -0.009 & 0.676 & 0.280 & -1.10 & 0.37 \\
  Prior $h=0.72 \pm 0.02$  &  196.6 & 0.006 & 0.710 & 0.254 & -1.15 & 0.81 \\
  Prior $h=0.62 \pm 0.02$   &  197.3 & -0.026 & 0.638 & 0.314 & -0.93 & -0.93 \\
\hline
  $\tilde{w}_0$ and $\tilde{w}_1$ ($z_\ast = 0.5$) & 195.5 & -0.010 & 0.676 & 0.280 & -1.07 & -0.99 \\
  Prior $h=0.72 \pm 0.02$ & 196.7 & 0.002 & 0.710 & 0.254 & -1.15 & -0.80 \\
  Prior $h=0.62 \pm 0.02$  & 197.3 & -0.025 & 0.638 & 0.315 & -0.94 & -1.28 \\
   \hline
 \end{tabular}
 \caption{Best fit values of $\Omega_k$ and $h$ for various assumptions on 
   dark energy. Here we also show the best fit 
   values for marginalized parameters 
   such as $\Omega_m, w_0 (\tilde{w}_0)$ and $w_1 (\tilde{w}_1)$. }
\end{table}

The final remark is that, in the previous works
\cite{Ichikawa:2006qb,Ichikawa:2006qt}, it has been shown that when we
parametrize dark energy equation of state as Eq.~\eqref{eq:eos2} and
take $z_\ast \sim 0.5$, the region of an open universe as large as
$\Omega_k \sim 0.2$ is allowed.  However, for the result presented in
Fig.~\ref{fig:Ok_h_margOm_w0}, this is not the case even though we use
the same type of parametrizaion.  This is because we in this paper use a strong
prior $w_X \le -1/3$ on the equation of state given in
Eq.~\eqref{eq:eos_prior} but in the previous works
\cite{Ichikawa:2006qb,Ichikawa:2006qt}, a weak prior $w_X \le 0$ has
been adopted\footnote{
  The method of this paper to obtain a constraint from cosmological
  data is also slightly different from that used in
  \cite{Ichikawa:2005nb,Ichikawa:2006qb,Ichikawa:2006qt}.  However,
  the difference in the method is irrelevant to the conclusion on the
  constraint on the curvature of the universe.
}.  As is already noticed in
Refs.~\cite{Ichikawa:2006qb,Ichikawa:2006qt}, an open universe tends
to be preferred for dark energy whose equation of state approaches to
zero at earlier time. Since we remove this possibility by adopting a strong
prior, we do not have a large allowed region with an open
universe. 
In other words, our result here confirmed explicitly that the dark
energy whose equation of state is close to zero at earlier time, which
is sometimes called early dark energy, can help to allow an open
universe at least as long as the background evolution is concerned.

\section{Conclusions and Discussion}

We studied the constraints on the Hubble constant from CMB, BAO and SN
assuming several types of dark energy parametrization.  Although the
Hubble constant and dark energy are both important issues in cosmology
today, these two subjects have not been investigated much
simultaneously.  First we investigated the constraints in the
$\Omega_m$--$h$ plane assuming several dark energy parametrizations.
The constraints on the Hubble constant from the combination of CMB,
BAO and SN observations obtained under the different dark energy
models and priors are summarized in Table~\ref{tab:margDE} when a flat
universe is assumed and in Table~\ref{tab:margDE_Ok} when the flatness
assumption is dropped respectively.  It is noticed that the
constraints are not affected drastically by the dark energy model
assumed and/or the assumption of the flatness of the universe.  It is
rather more affected by the choice of the the SN data set: the Gold06
set gives slightly lower value of $h$ than the Davis07 set.  We can
conservatively conclude that $H_0 < 59$ and $H_0 > 76$ are highly
unlikely from these cosmological observations: these parameter regions
are not allowed at 2$\sigma$ level for any dark energy parametrization
even if we do not restrict ourselves to a flat universe. Since the 
distance ladder estimations of $H_0$ have somewhat large systematic
errors at present, we are not at the stage of arguing any possible
discrepancies among these measurements now.  Nevertheless, it is worth
mentioning in passing that the Sandage's central value $H_0 = 62$
\cite{Sandage:2006cv} and the Macri's value $H_0 = 74$
\cite{Macri:2006wm} are fairly close to the limit we have obtained
here.

We have also investigated the constraints on some dark energy
parameters assuming several priors on the Hubble constant.  The
constraints are derived with and without the assumption of the
flatness of the universe.  Using the present cosmological observations
assuming no Hubble prior, we have found that a cosmological constant
and a flat universe can fit all the data satisfactorily.  When we
impose a prior on the Hubble constant, we have adopted Gaussian
priors $h=0.72 \pm 0.02$ and $h= 0.62 \pm 0.02$.  The central values
are those of Freedman's \cite{Freedman:2000cf} and Sandage's
\cite{Sandage:2006cv} but the error  are taken to be a
hypothetical value to give a meaningful effect on the dark energy
parameter estimation.  We have found that, even with some
limited options discussed in this paper for the assumptions with respect
to the prior on the Hubble constant, a parametrization of the dark
energy equation of state, the curvature of the universe, they can have
a great influence on determining the nature of dark energy.  It should
also be mentioned that the choice of the SN data set affects the allowed
region. 
We demonstrate these points by taking a cosmological constant as
a reference dark energy model because it is the simplest and most conventional
model which can fit the observational data sets.
For example, when we adopt the
parametrization Eq.~\eqref{eq:eos} in a flat universe, the prior $h=
0.62 \pm 0.02$ reject a cosmological constant at 2$\sigma$ level for
the Gold06 data set, but the allowed region broadens to allow a
cosmological constant if we do not adopt the flatness assumption or if
we instead adopt the prior $h= 0.72 \pm 0.02$. Moreover, if we adopt
the parametrization Eq.~\eqref{eq:eos2} with $z_\ast = 0.5$ and the
prior $h= 0.72 \pm 0.02$, the Gold06 data set is in good agreement
with a cosmological constant in a flat universe, but if we drop the
flatness condition and marginalize over the curvature, a cosmological
constant is disfavored at 2$\sigma$ level.
These examples imply that our understanding of the nature 
of dark energy can be varied by the assumptions on the Hubble constant,
a parametrization of dark energy equation of state and the curvature of the universe. 

Finally, we have investigated the constraints on the curvature of the
universe assuming several types of dark energy and the priors on the
Hubble constant.  In contrast to the constraints on the Hubble
constant, we see the result depends on the dark energy parametrization
we adopt. For the cases with a cosmological constant and a constant
equation of state, the allowed region occupies larger area in a closed
universe, whereas for the case with a time-varying equation of state
parametrized as Eq.~\eqref{eq:eos2} with $z_\ast = 0.5$, the allowed
region extends into a region of an open universe.  Since there is an
obvious positive correlation between $h$ and $\Omega_k$, the prior $h=
0.62 \pm 0.02$ exaggerates the preference for a closed universe and
the prior $h= 0.72 \pm 0.02$ for an open universe.  This is what we
see in Fig.~\ref{fig:Ok_h_margOm_w0}.  We also reconfirmed that the
preference of an open universe is caused by the equation of state for
dark energy which is close to $0$ at earlier time.  This is because,
we have found that $\Omega_k$ as large as 0.2 is allowed in our
previous papers Refs.~\cite{Ichikawa:2006qb,Ichikawa:2006qt} under the
weak prior $w_X \le 0$, whereas $\Omega_k$ is found to be well below
0.1 in this paper under the stronger prior $w_X \le -1/3$.

Since dark energy is one of the most important problems in science
today, a large number of works are focusing on dark energy
itself. However, when one tries to probe the nature of dark energy,
other cosmological parameters such as the Hubble constant, which was discussed in
this paper, should necessarily be involved in various manners.  In
light of precise measurements of cosmology that we are having now, the
works from this kind of viewpoint should be done to check our
understanding of cosmology and may also give insights to probe the
present state and the evolution of the universe.

\bigskip
\bigskip

\noindent {\bf Acknowledgments:} This work was supported in part by
the Grant-in-Aid for Scientific Research from the Ministry of
Education, Science, Sports, and Culture of Japan, No.\,18840010 (K.I.)
and No.\,19740145 (T.T.).

\bigskip
\bigskip

\pagebreak

\appendix

\noindent {\bf \Large Appendix}

\section{Effects of dark energy perturbation on CMB} \label{sec:DE_pert}

In the analysis for the constraints, we make use of the quantities
which can be thoroughly determined by the background evolution. Thus
we only take into account the the modification to the background
evolution by dark energy to obtain constraints on some parameters
although dark energy component can fluctuate in general to affect the
comic density fluctuation such as CMB anisotropies.  As mentioned in
the text, when dark energy becomes dominant component at late time,
the effect of fluctuation is not significant except on large scales
and that from the modification to the background is enough to extract
the cosmological constraints.  However, when the equation of state for
dark energy approaches to zero at earlier time, which means that the
energy density of dark energy behaves $\rho_X \propto a^{-3}$ and can
be comparable to that of matter, fluctuation of dark energy becomes
important to affect the structure of acoustic peaks. In such a case,
fluctuation of dark energy should be properly taken into account in
the analysis. Thus, as for the CMB, the constraint from the distance
measure such as the acoustic scale which is given by the information
on the background evolution alone becomes invalid.  This is one of the
reason why we assume the prior of Eq.~\eqref{eq:eos_prior} 
to avoid such a case where dark energy has
significant fraction at earlier time.  In addition, when fluctuation
of dark energy becomes important, other nature of dark energy such as
the effective sound speed, which we denote here as $c_X^2$, can
also modify density fluctuation. In this sense, the equation of state
is not enough to consider the effect of dark energy.

\begin{figure}[t]
\begin{center}
\scalebox{0.62}{\includegraphics{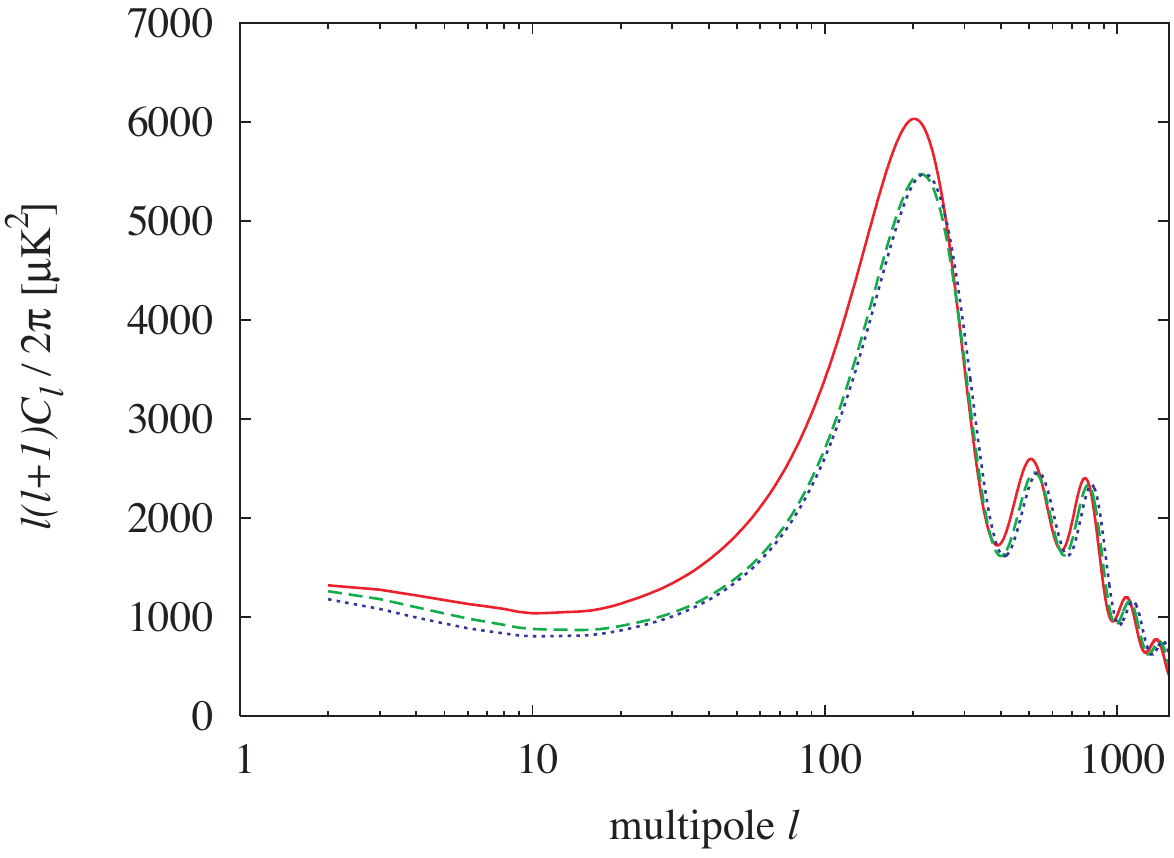}}
\hfill
\scalebox{0.62}{\includegraphics{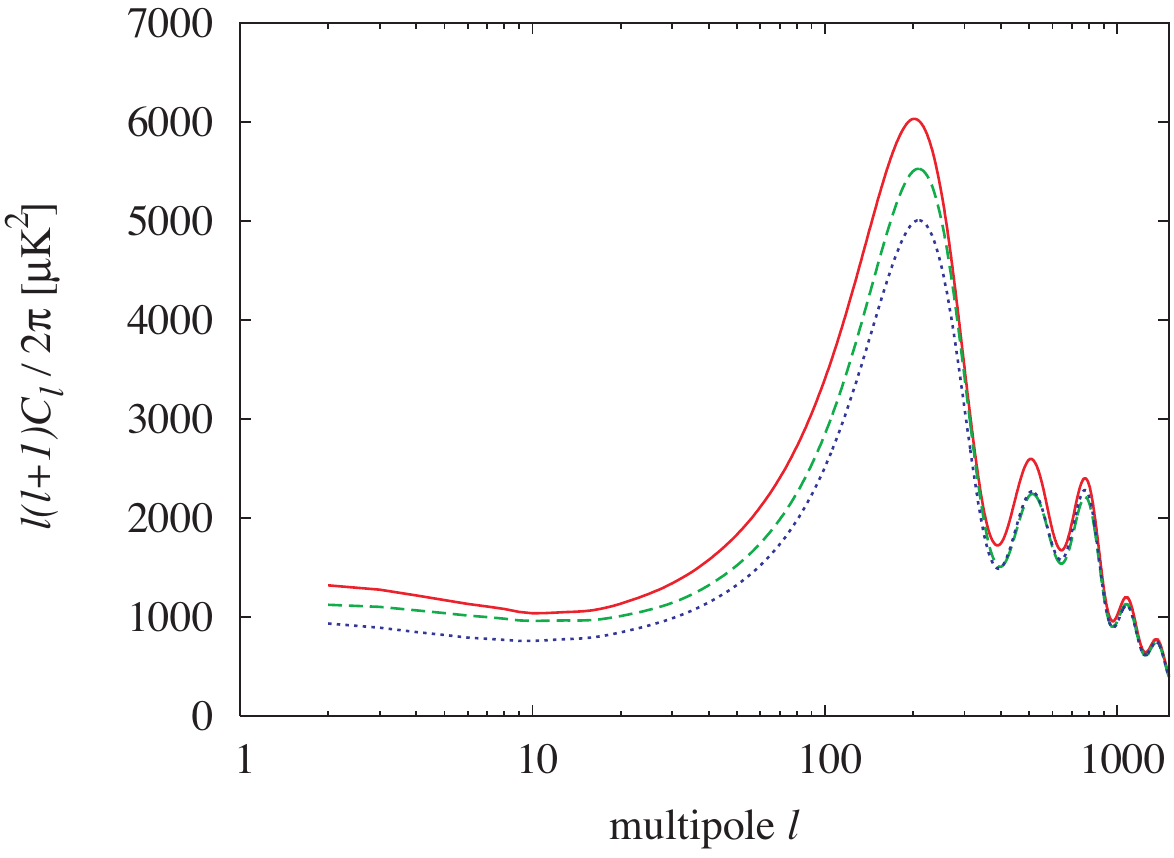}}
\caption{(Left panel) CMB power spectra for a dark energy model with the
  parametrization of Eq.~\eqref{eq:eos}.  Here we fix the value of
  $w_0 = -1$ and take several values for $w_1$ as $w_1 = 1$ (red solid
  line), $0.6$ (green dashed line) and $0$ (blue dotted line).  In
  this panel, we assume the effective sound speed as $c_X^2=1$.
  (Right panel) CMB power spectra for the equation of state $w_0 = -1 +
  (1-a)$ (i.e., we take $(w_0, w_1) = (-1, 1)$ for the parametrization
  of Eq.~\eqref{eq:eos}. ) The cases with $c_X^2 = 1$ (red solid
  line), $0.1$ (green dashed line) and $0$ (blue dotted line) are
  shown. }
\label{fig:cl}
\end{center}
\end{figure}

To see how the nature of dark energy can affect CMB, we show the CMB
power spectra in Fig.~\ref{fig:cl} for several cases of the equation
of state and the effective sound speed. Here we assume the
parametrization of Eq.~\eqref{eq:eos} for dark energy \footnote{
Other parametrization which makes dark energy density non-negligible at the epoch 
of recombination is adopted in Ref.~\cite{Caldwell:2003vp,Doran:2006kp} and constraints on 
the dark energy parameters are investigated. 
}.
 In the left panel, we fix the value of $w_0$ as $w_0=-1$ and vary $w_1$ as $w_1 =
1$ (red solid line), $0.8$ (green dashed line) and $0$ (blue dotted
line).  For other cosmological parameters, we assume the mean values
of a power-law $\Lambda$CDM model from WMAP3 alone analysis as
$\omega_m =0.1277, \omega_b =0.02229, \tau =0.089, h =
0.732$ and $n_s = 0.958$.  As seen from the figure, when we compare
the cases with $w_1=0$ and $0.6$, the power spectra is just shifted to
smaller $l$ but the structure of acoustic peaks is unchanged. This
shift is caused by the the change of the angular diameter distance to
last scattering surface due to the modification to the background
evolution.  However, when we take $w_1=1$ in which energy density of
dark energy is not negligible compared to that of matter at earlier
time, the structure of acoustic peaks is significantly modified
because fluctuation of dark energy can affect it in addition to the
shift of the peaks in this case.
Furthermore, this change depends on
the nature of dark energy fluctuation. In the right panel of
Fig.~\ref{fig:cl}, we plot the cases with $c_X^2 = 1$ (red solid
line), $0.1$ (green dashed line) and $0$ (blue dotted line) for a
parameter set $(w_0, w_1) =(-1,1)$.  Even though the equation of state
is the same (namely, the background evolution is the same), the CMB
power spectra are different when one assumes different sound speed.  In
this kind of case,  since the effective sound speed (and/or possibly another
property of dark energy fluctuation) can affect the CMB,
one should take into account the whole information
of CMB power spectrum and constraints from the background evolution 
become invalid. 

\begin{figure}
\begin{center}
\scalebox{0.58}{\includegraphics{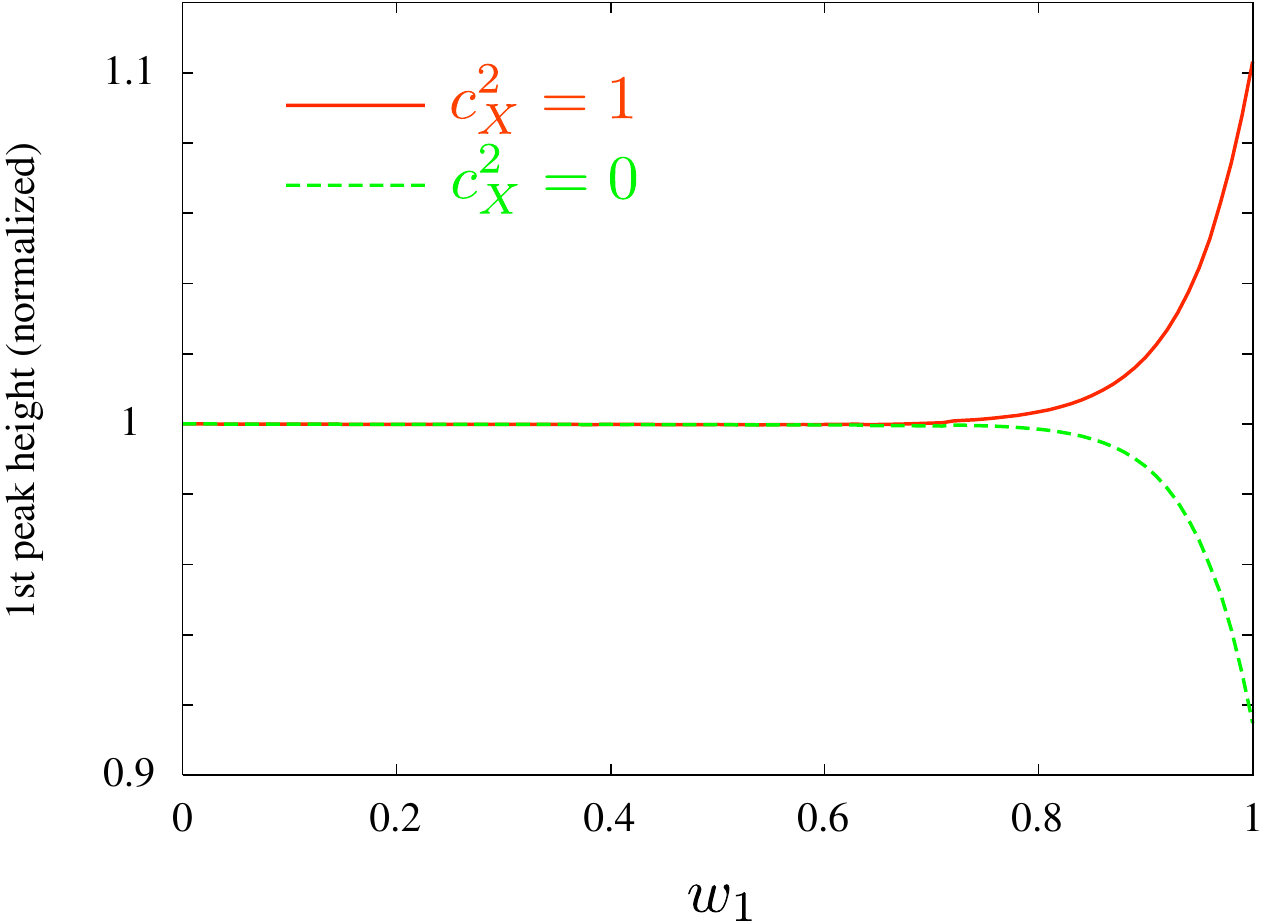}}
\hfill
\scalebox{0.58}{\includegraphics{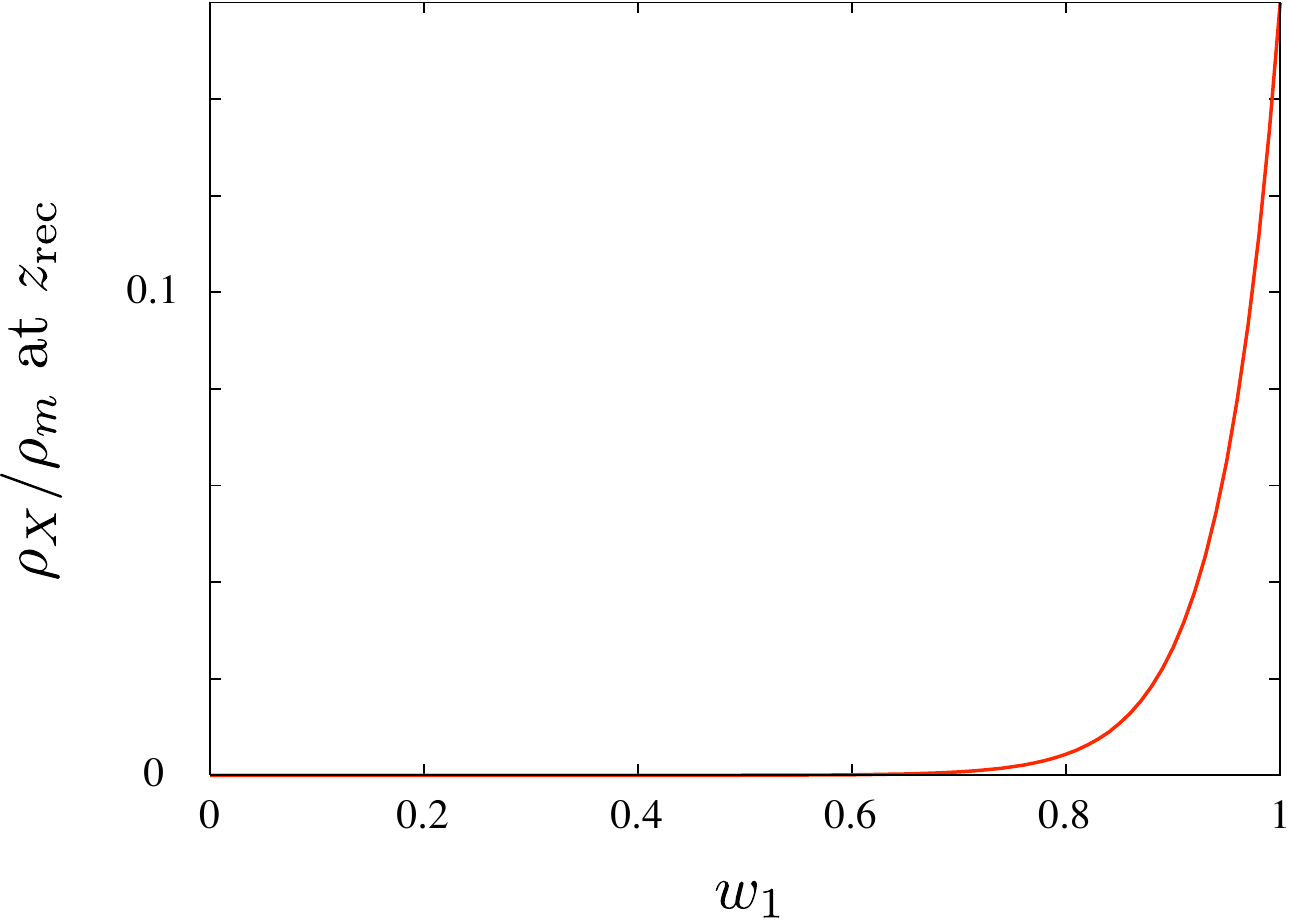}}
\caption{(Left panel) The height of the first peak as a function of $w_1$ of the
  parametrization Eq.~\eqref{eq:eos} for the cases with $c_X^2 = 1$
  (red solid line) and $0$ (green dashed line).  The value of $w_0$
  is fixed as $w_0=-1$.  (Right panel) The ratio of the energy density of dark energy to that
  of matter at $z=1089$ as a function of $w_1$.
  }
\label{fig:ratio}
\end{center}
\end{figure}

We made another plot, Fig.~\ref{fig:ratio}, to see when
neglecting the fluctuation of dark energy component is valid. 
In the left panel, we plot the height of the CMB first peak as a function of $w_1$ for the
parametrization Eq.~\eqref{eq:eos} with fixing $w_0= -1$. 
We can see that it is unchanged for $w_1 \lesssim 0.7$, which indicates
that the fluctuation of dark energy does not affect the structure of the acoustic peaks.
In the right panel, as a function of $w_1$, we plot
the ratio of energy densities of dark energy and matter at the
recombination epoch, $z_{\rm rec} = 1089$.  
Notice that when $w_1 \lesssim 0.7$, energy
density of dark energy is negligible compared to that of matter at
recombination epoch.  In other words, the effect of fluctuation of
dark energy can be neglected when energy density of dark energy is
small enough at earlier time.
In such a case, since the information on
the background evolution alone well captures the effect of dark energy,
we can constrain dark energy parameters by only studying the shift of acoustic
peaks. 
Thus, our method to obtain constraints from observations of CMB
is justified when the equation of state is in a range where the energy
density of dark energy is negligible at earlier time. 
The prior we take in this paper, Eq.~\eqref{eq:eos_prior} $w_X \le -1/3$, can 
satisfy this requirement.
The final comment is that, as can be inferred from Fig.~\ref{fig:ratio}, a slightly looser 
prior like $w_X \le -0.3$ may make dark energy subdominant at the epoch of recombination 
and justify our analysis (and would not change our results much).
However, since this condition depends on other cosmological parameters such as $\Omega_X$ and $\Omega_m$,
we adopt the conservative prior of $w_X \le -1/3$.

\section{How $\Omega_m$ and $\Omega_k$ are determined from CMB, BAO and SN} \label{sec:OmOk}
In Sec.~\ref{subsec:constraint_H}, we found that $h$ is constrained
rather tightly regardless of the dark energy model and/or the
assumption of the flatness of the universe if all the CMB, BAO and SN
data are combined. We also argued that this is equivalent to the
determination of $\Omega_m$ from the acoustic scales, $\theta_A$ and
$D_V(0.35)$, and the luminosity distance $d_L$ since $\omega_m =
\Omega_m h^2$ is given by the height of the CMB acoustic peak.  Thus,
in this appendix, we explain how these three types of cosmological
data sets can determine $\Omega_m$ without referring to any particular
dark energy model and without assuming the flat universe.

\subsection{What SN and CMB determine} \label{sec:OmOk_SNCMB} 
\begin{figure}[t]
    \centerline{\scalebox{1}{\includegraphics{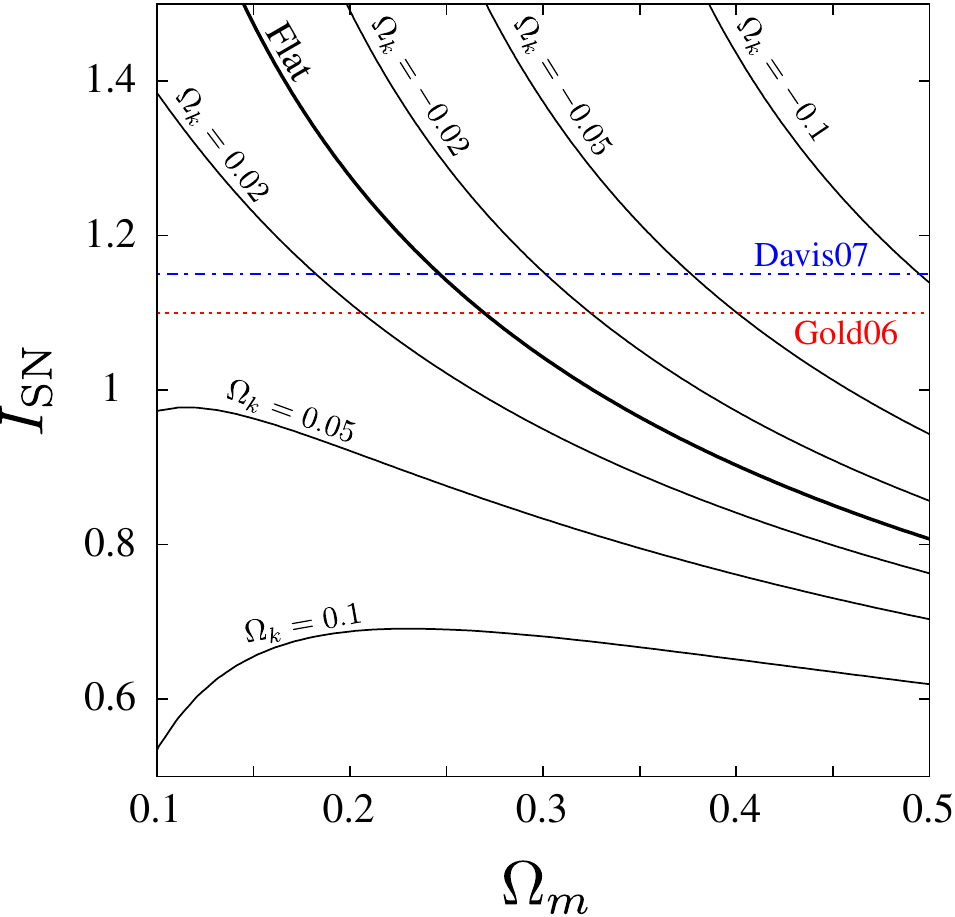}}}
    \caption{Contours of $r_\theta(z_{\rm rec}) = 14.3$\,Gpc for
      several values of $\Omega_k$ (black solid lines) and $I_{\rm
        SN}$ derived from observations of Gold06 (red dotted line) and
      Davis07 (blue dot-dashed line). }
    \label{fig:Om_Isn}
\end{figure}

In this
section, we focus on SN and CMB observations.  We start with defining
\begin{eqnarray}
I_{\rm SN} \equiv \int_0^{z_{\rm SN}} \frac{dz^\prime}{H(z^\prime)/H_0}, \label{eq:Isn}
\end{eqnarray}
where $z_{\rm SN}$ denotes the highest redshift to which SN data can
probe.  For the present data, we take $z_{\rm SN}=1.8$. Using this
integral, the comoving angular distance to the last scattering surface
Eq.~(\ref{eq:r_theta}) is given by
\begin{eqnarray}
r_\theta(z_{\rm rec}) 
&=& \frac{1}{H_0 \sqrt{|\Omega_k|}}
\mathcal{S}  
\left( \sqrt{|\Omega_k|} \left\{ I_{\rm SN} + \int_{z_{\rm SN}}^{z_{\rm rec}} \frac{dz'}{H(z')/H_0} \right\} \right) \\
&\approx& \frac{3.0 \times 10^3\, {\rm Mpc}}{\sqrt{\omega_m}} \frac{\sqrt{\Omega_m}}{\sqrt{|\Omega_k|}} \mathcal{S}   \left( \sqrt{|\Omega_k|} \left\{ I_{\rm 
SN} + 2 \Omega_m^{-1/2} (1+z_{\rm SN})^{-1/2} \right\} \right).  \label{eq:rth_app}
\end{eqnarray}
In the second line, we analytically performed the integration by
approximating the universe to be matter-dominated for $z_{\rm SN} < z
< z_{\rm rec}$ and neglected the term proportional to $(1+z_{\rm
  rec})^{-1/2}$ since $z_{\rm SN} \ll z_{\rm rec}$. Now, if $I_{\rm
  SN}$ is known from SN observation, since CMB observation gives
$\omega_m$ and $r_\theta(z_{\rm rec})$, we obtain the degeneracy
relation between $\Omega_m$ and $\Omega_k$. This is drawn in
Fig.~\ref{fig:Om_Isn} as a contour of $r_\theta(z_{\rm rec}) =
14.3$\,Gpc (this is derived from Eq.~(\ref{eq:theta_A}) and
$r_s(z_{\rm rec}) = 149$\,Mpc, which is in turn from
Eq.~(\ref{eq:sound_horizon}) using $\omega_m = 0.1277$ and $\omega_b =
0.02229$) in the $\Omega_m$--$I_{\rm SN}$ plane for several values of
$\Omega_k$. Specifically, if we impose a flat universe prior,
$\Omega_k = 0$, $\Omega_m$ is determined uniquely. In
Ref.~\cite{Knox:2005hx}, essentially the same argument is used to
investigate the prospect of measuring $\Omega_k$ very precisely.

The next issue is how tightly $I_{\rm SN}$ is constrained by the
data. The SN data such as Gold06 or Davis07 used in this paper consist
of the $z$--$\mu$ relation with an error on $\mu$ where $\mu$ is given by
Eq.~(\ref{eq:mu}).  The integral $I_{\rm SN}$ is directly obtained
from $\mu(z_{\rm SN})$ if $H_0$ and $M$ are known.  However, the
uncertainties in them cannot be resolved by SN data alone. Thus to
cancel these uncertainties, we use $\mu$ at another redshift $z_1$
which is close to zero. ($z_1 \ll 1$ is required not to make
$\mu(z_1)$ dependent on cosmology. Notice that, as we will adopt
below, approximately we can use the relation $d_L(z) \simeq z/H_0$ for
small $z$.) To cancel the constants, we take a difference of $\mu(z)$ as
\begin{eqnarray}
\mu(z_{\rm SN}) - \mu(z_1) = 5 \log \frac{d_L(z_{\rm SN})}{d_L(z_1)}.
\end{eqnarray}
Since 
\begin{eqnarray}
d_L(z_{\rm SN}) = \frac{1+z_{\rm SN}}{H_0 \sqrt{|\Omega_k|} }
\mathcal{S}  \left( \sqrt{|\Omega_k|}\, I_{\rm SN} \right),
\end{eqnarray}
and, $d_L(z_1) \approx z_1/H_0$ for $z_1 \ll 1$, we obtain
\begin{eqnarray}
I_{\rm SN} = \frac{1}{\sqrt{|\Omega_k|}} \mathcal{S}^{-1} \left( \frac{z_1 \sqrt{|\Omega_k|} }{1+z_{\rm SN}} 10^{(\mu(z_{\rm SN}) - \mu(z_1))/5}  \right).
\end{eqnarray}

We derive $\mu(z_{\rm SN})$ and $\mu(z_1)$ by fitting the SN data to a
power-law function $\mu(z) = a z^p$. For the Gold06 set,
\begin{eqnarray}
a_G &=& 44.288 \pm 0.022, \\
p_G &=& (6.048 \pm 0.033) \times 10^{-2},
\end{eqnarray}
and for the Davis07 set, 
\begin{eqnarray}
a_D &=& 44.278 \pm 0.023, \\
p_D &=& (6.123 \pm 0.030) \times 10^{-2}.
\end{eqnarray}
Note that it does not refer to a specific parametrization for dark
energy. The $\chi^2$'s are 158.9 for the Gold06 data and 199.0
for the Davis07 data which are comparable to the values by the usual
approach \cite{Ichikawa:2006qt,Barger:2006vc} of assuming some dark
energy models.  $I_{\rm SN}$ has an uncertainty as regards the value
of $z_1$. However, a numerical experiment reveals that $I_{\rm SN}$
barely depends on the value of $z_1$ for $0.02 \lesssim z_1 \lesssim
0.05$ ($z \sim 0.02$ is the lowest redshift for the SN data we here
use). Thus, we take $z_1 = 0.05$ hereafter.  Then in a flat
universe,
\begin{eqnarray}
I_{\rm SN,G} &=& 1.10, \\
I_{\rm SN,D} &=& 1.15,
\end{eqnarray}
respectively for Gold06 and Davis07.  Notice that they do not depend
much on the curvature. In fact, they change less than 1\% for
$|\Omega_k| < 0.2$.  These values are plotted as horizontal lines in
Fig.~\ref{fig:Om_Isn}. They cross the contour which satisfies the CMB
observation for a flat universe (denoted by the black thick solid line) at around $\Omega_m = 0.25$. This is
consistent with the full analysis result as presented in
Sec.~\ref{subsec:constraint_H}. Moreover, we can see from the figure
that since $I_{\rm SN,G} < I_{\rm SN,D}$ (which in turn comes from the
difference in the power-law slope $p_G < p_D$), Gold06 favors larger
$\Omega_m$ than Davis07 does. This is another point which we have
noted in Sec.~\ref{subsec:constraint_H}. We believe that these
explanations are independent of the dark energy model we adopt because
we use the SN data without referring to dark energy.

However, Fig.~\ref{fig:Om_Isn} also shows that if we abandon the
flatness assumption and shift $\Omega_k$ from zero, different value of
$\Omega_m$ is favored. This is why we cannot constrain $\Omega_m$ much
from CMB+SN when we marginalize over $\Omega_k$.

\subsection{What BAO adds} \label{sec:OmOk_BAO}
The BAO data gives $D_V(z_{\rm BAO})$ defined by Eq.~(\ref{eq:DV})
where $z_{\rm BAO} = 0.35$. This is written as
\begin{eqnarray}
  D_V(z_{\rm BAO})^3
  &=& r_\theta(z_{\rm BAO})^2 H_0^{-1} \frac{z_{\rm BAO}}{\sqrt{\Omega_m (1+z_{\rm BAO})^3 +\Omega_k (1+z_{\rm BAO})^2 +(1-\Omega_m -\Omega_k) f
(z_{\rm BAO})}}, \nonumber \\
\end{eqnarray}
where $f(z)$ is a function which expresses the evolution of the dark
energy density.  Here, the dark energy dependence enters in two
places, $r_\theta(z_{\rm BAO})$ and $f(z_{\rm BAO})$. Since $z_{\rm
  BAO}$ is relatively small, we may neglect the dark energy evolution
(namely, we approximate dark energy as a cosmological constant)
to approximate as $f(z_{\rm BAO}) \approx 1$.  For $r_\theta(z_{\rm
  BAO})$, since this is written as
\begin{eqnarray}
r_\theta(z_{\rm BAO}) 
&=&\frac{1}{H_0 \sqrt{|\Omega_k|} }
\mathcal{S}  
\left( \sqrt{|\Omega_k|} \, I_{\rm BAO} \right),
\end{eqnarray}
where $I_{\rm BAO}$ is the integral similar to Eq.~(\ref{eq:Isn}) with
$z_{\rm SN}$ replaced by $z_{\rm BAO}$, it can be calculated without
referring to a dark energy model provided that we combine with the SN data. This
is because we can infer $I_{\rm BAO}$ by using the fit to the SN data
in the same manner to derive $I_{\rm SN}$ in Appendix \ref{sec:OmOk_SNCMB}.
Similarly to $I_{\rm SN}$, $I_{\rm BAO}$ depends slightly on
the SN data (but not on $z_1$ or the curvature). Gold06 and Davis07
give respectively
\begin{eqnarray}
I_{\rm BAO, G} &=& 0.31, \\
I_{\rm BAO, D} &=& 0.32.
\end{eqnarray}
Now, if we fix $\omega_m$ using the CMB value, we have a relation
between $\Omega_m$ and $\Omega_k$ determined from SN and BAO data
for a measured $D_V(0.35)$.

Thus, here and in Appendix \ref{sec:OmOk_SNCMB}, we have 
replaced the dark energy dependent part of CMB and BAO observables
($r_\theta(z_{\rm rec})$ and $D_V(0.35)$) by the empirical values
($I_{\rm SN}$ and $I_{\rm BAO}$) inferred from the SN data.  Namely,
we can now draw contours of $r_\theta(z_{\rm rec})$ and $D_V(0.35)$ in
the $\Omega_m$--$\Omega_k$ plane without referring to any dark energy
model.  In Fig~\ref{fig:Ok_Om}, we draw the contours $r_\theta(z_{\rm
  rec})=14.3$\,Gpc (see Appendix \ref{sec:OmOk_SNCMB}) by the black
solid line and $D_V(0.35)=1402$\,Mpc (see Sec.~\ref{sec:method_BAO})
by the black dotted line. We have fixed $\omega_m$ to the CMB value of
0.1277. Since the two SN data sets give slightly different values of
$I_{\rm SN}$ and $I_{\rm BAO}$ as mentioned above, we show the results
in two panels separately (the Gold06 data are used in the left panel
and the Davis07 data in the right panel). The $r_\theta(z_{\rm rec})$
contours run in somewhat diagonal direction, showing the degeneracy
between $\Omega_m$ and $\Omega_k$ for the CMB+SN combination as
mentioned in Appendix \ref{sec:OmOk_SNCMB}.  In contrast, the $D_V(0.35)$
contours run almost horizontally, showing that the BAO+SN combination
is insensitive to $\Omega_k$ and can determine $\Omega_m$ regardless
of the assumption on the curvature of the universe. This is reasonable
because BAO measures the distance to relatively low redshift
($z=0.35$). Ref.~\cite{Eisenstein:2005su} has derived a linearized
relation from BAO measurement: $\Omega_m = 0.273 + 0.137 \Omega_k$
(Eq.~(6) in Ref.~\cite{Eisenstein:2005su}), which is quite similar to
our slope for the $D_V(0.35)$ contours. But also note that this
relation has been derived only for a cosmological constant.  If we
consider the case with a constant equation of state or a time-varying
equation of state, $\Omega_m$ cannot be determined from BAO alone.
Our point is that when BAO and SN are combined, $\Omega_m$ can be
determined regardless of the assumption on the curvature of the
universe and dark energy model.

In order to check the validity of these approximations, we plot the
allowed regions by the full analysis such as explained in
Sec.~\ref{sec:method}. The $r_\theta(z_{\rm rec})$ contours are
compared with the analysis using $\chi^2$ constructed from $\theta_A$,
$\omega_m$ and the SN data set. The $D_V(0.35)$ contours are compared
with the one using $\chi^2$ from $D_V(0.35)$, $\omega_m$ and SN. For
completeness, we also performed the all-combined analysis using
$\theta_A$, $D_V(0.35)$, $\omega_m$ and SN, which can be compared with
the analyses of $\Omega_m$ and $\Omega_k$ done in
Sec.~\ref{subsec:constraint_H} and
Sec.~\ref{subsec:constraint_Ok}. The $\chi^2$'s are minimized over $h$
and dark energy parameters. We analyzed several dark energy models
parametrized as Eqs.~(\ref{eq:eos}) and (\ref{eq:eos2}) but the
results do not show much difference. In Fig.~\ref{fig:Ok_Om}, we show
2$\sigma$ allowed regions for the cases with Eq.~(\ref{eq:eos}) (blue
dot-dashed lines) and Eq.~(\ref{eq:eos2}) with $z_\ast = 0.5$ (red
dashed lines) as examples.  Note that the allowed regions for two
models look alike. We can also see that the regions basically extend
in the directions of the approximate $r_\theta(z_{\rm rec})$ and
$D_V(0.35)$ contours. Therefore, it is considered to be appropriate to drop the dark energy
dependence by using the SN data as we have done here.

 \begin{figure}[htb]
    \centerline{\scalebox{1}{\includegraphics{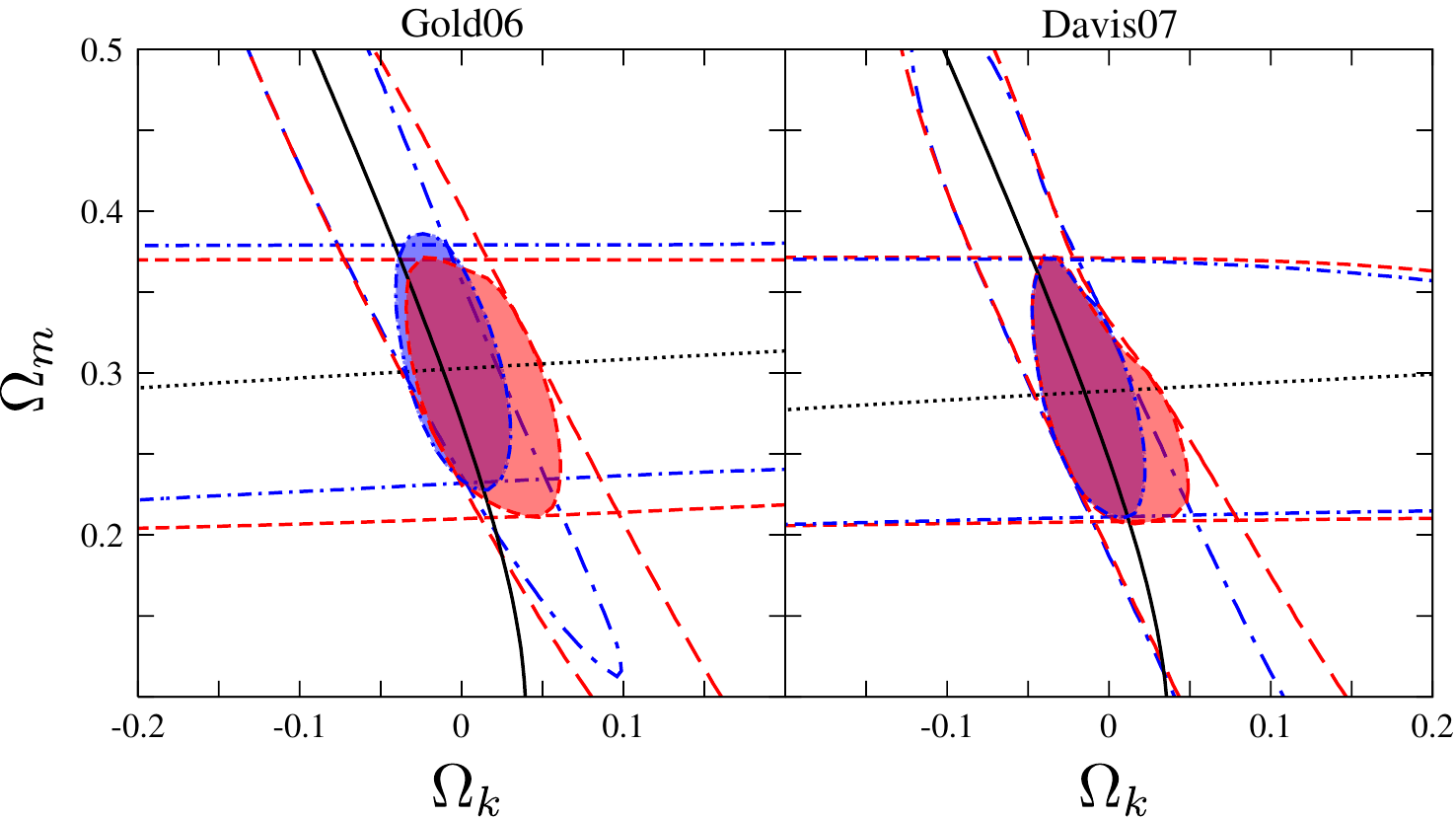}}}
    \caption{Contours of $r_\theta(z_{\rm rec})=14308.5$\,Mpc
      (Eq.~(\ref{eq:rth_app})) are drawn in black solid lines and
      those of $D_V(0.35)=1402$\,Mpc are drawn in black dotted lines.
      Here, $r_\theta(z_{\rm rec})$ and $D_V(0.35)$ are approximate
      expressions in which dependence on dark energy model is dropped
      by using empirical values from SN data as discussed in
      Appendix \ref{sec:OmOk_SNCMB} and \ref{sec:OmOk_BAO}.  Also shown
      are 2$\sigma$ allowed regions derived by the full analysis
      assuming some dark energy models and marginalizing over two dark
      energy parameters and $h$. The models are: (i)
      the parametrization Eq.~\eqref{eq:eos} and (ii) the parametrization Eq.~\eqref{eq:eos2} with
      $z_\ast=0.5$. The bands running diagonally are from SN,
      $\theta_A$ and $\omega_m$ (blue dot-long-dashed line for Model
      (i) and red long-dashed line for Model (ii)), and the horizontal
      bands are from SN, $D_V(0.35)$ and $\omega_m$ (blue
      dot-short-dashed line for Model (i) and red short-dashed line
      for Model (ii)). The shaded regions are from all data combined,
      SN, $\theta_A$, $D_V(0.35)$ and $\omega_m$ (the regions are
      bounded by blue dot-dashed line for Model (i) and red dashed
      line for Model (ii)). We can see that the contours of the
      approximate expressions well indicate the degeneracy directions
      of the corresponding data combinations. We use the Gold06 SN
      data set in the left panel and the Davis07 data in the right
      panel.}
    \label{fig:Ok_Om}
\end{figure}

\end{document}